\documentclass[a4paper,fleqn,usenatbib]{mnras}

\usepackage{newtxtext,newtxmath}

\usepackage[T1]{fontenc}
\usepackage{ae,aecompl}

\usepackage{graphicx}	
\usepackage{amsmath}	
\usepackage{amssymb}	

\usepackage{graphicx}
\usepackage{epsfig}
\usepackage{multirow}
\usepackage{float}
\usepackage{amsmath}
\def\mnras{MNRAS}
\def\apj{ApJ}
\def\aj{AJ}
\def\apjl{ApJL}
\def\apjs{ApJS}
\def\aap{A\& A}

\def\pasj{PASJ}
\def\nat{Nature}
\def\axj{AX J1745.6-2901}
\def\sgra{Sgr A$^\star$}
\def\xmm{{\it XMM-Newton}}
\def\cxo{{\it Chandra}}
\def\swift{{\it Swift}}

\title[Dust Scattering Halo around AX J1745.6-2901]{Probing the Interstellar Dust towards the Galactic Centre: Dust Scattering Halo around  AX J1745.6-2901}

\author[C. Jin, et al.]{
Chichuan Jin$^{1}$\thanks{E-mail: chichuan@mpe.mpg.de},
Gabriele Ponti$^{1}$,
Frank Haberl$^{1}$,
Randall Smith$^{2}$
\\
$^{1}$Max-Planck-Institut f\"{u}r Extraterrestrische Physik, Giessenbachstrasse, D-85748 Garching, Germany\\
$^{2}$Smithsonian Astrophysical Observatory, 60 Garden Street, Cambridge, MA 02138, USA\\
}

\date{prepared for MNRAS}

\pubyear{2017}

\begin{document}
\label{firstpage}
\pagerange{\pageref{firstpage}--\pageref{lastpage}}
\maketitle

\begin{abstract}
\axj\ is an X-ray binary located at only 1.45 arcmin from Sgr A$^\star$, showcasing a strong X-ray dust scattering halo. We combine \cxo\ and \xmm\ observations to study the halo around this X-ray binary. Our study shows two major thick dust layers along the line of sight (LOS) towards \axj. The LOS position and $N_{H}$ of these two layers depend on the dust grain models with different grain size distribution and abundances. But for all the 19 dust grain models considered, dust Layer-1 is consistently found to be within a fractional distance of 0.11 (mean value: 0.05) to \axj\ and contains only (19-34)\% (mean value: 26\%) of the total LOS dust. The remaining dust is contained in Layer-2, which is distributed from the Earth up to a mean fractional distance of 0.64. A significant separation between the two layers is found for all the dust grain models, with a mean fractional distance of 0.31. Besides, an extended wing component is discovered in the halo, which implies a higher fraction of dust grains with typical sizes $\lesssim$ 590 \AA\ than considered in current dust grain models. Assuming \axj\ is 8 kpc away, dust Layer-2 would be located in the Galactic disk several kpc away from the Galactic Centre (GC). The dust scattering halo biases the observed spectrum of \axj\ severely in both spectral shape and flux, and also introduces a strong dependence on the size of the instrumental point spread function and the source extraction region. We build {\sc Xspec} models to account for this spectral bias, which allow us to recover the intrinsic spectrum of \axj\ free from dust scattering opacity. If dust Layer-2 also intervenes along the LOS to \sgra\ and other nearby GC sources, a significant spectral correction for the dust scattering opacity would be necessary for all these GC sources.
\end{abstract}

\begin{keywords}
dust, extinction - X-rays: ISM; neutron stars:  X-rays: binaries: individual: \axj
\end{keywords}

\section{Introduction}
\label{sec-intro}
The centre of the Milky Way (i.e. the Galactic Centre, hereafter: GC) is one of the most attractive regions for astrophysical studies. It has been observed that thousands of point-like X-ray sources and extended sources are sitting in this region (e.g. Wang, Gotthelf \& Lang 2002; Muno et al. 2003, 2005, 2009; Degenaar et al. 2012, 2015; Ponti et al. 2015a, 2016), including Sgr A$^\star$ which is the super-massive black hole (SMBH) closest to Earth (e.g. Genzel, Eisenhauer \& Gillessen 2010). The inner few hundred parsecs contain a large concentration of molecular material (e.g. Morris 1996), which is revealing traces of previous active periods of Sgr A$^\star$ (see Ponti et al. 2013 for a review). However, the GC region is also highly extincted from the optical to the soft X-ray band (e.g. Becklin \& Neugebauer 1968; Fritz et al. 2011), elevating the importance of understanding the properties of gas and dust in front of the GC.

Previous studies about the diffuse emission and point sources around the GC LOS often suggest a large hydrogen column density of $N_{H}~\gtrsim10^{22}cm^{-2}$, as measured by absorption models in the X-ray spectral fitting (hereafter: $N_{H,abs}$; e.g. Terrier et al. 2010; Molinari et al. 2011; Degenaar et al. 2012). The large $N_{H,abs}$ also implies a large dust column assuming typical gas-to-dust ratios (e.g. Predehl \& Schmitt 1995). Recently, several papers show that dust scattering can have a significant impact on the observed X-ray spectra (Smith, Valencic \& Corrales 2016), causing significant bias to $N_{H,abs}$ if the dust scattering opacity is not properly considered in the spectral fitting (Corrales et al. 2016). Therefore, it is especially important to consider the effects of dust scattering when studying X-ray sources in the GC direction.

\subsection{Dust Scattering Halo in X-rays}
The idea that dust grains along the LOS can produce an observable halo around X-ray sources, the so called `dust scattering halo', was proposed fifty years ago (Overbeck 1965; Tr\"{u}mper \& Sch\"{o}nfelder 1973) and was first observed by the {\it Einstein} satellite around GX 339-4 (Rolf 1983). The investigation of the dust scattering halo can provide important information about the location of the dust along the LOS as well as the size and composition of the dust grains (e.g. Predehl et al. 1992; Predehl \& Schmitt 1995; Xiang, Zhang \& Yao 2005; Valencic \& Smith 2015; Heinz et al. 2016; Vasilopoulos \& Petropoulou 2016).

\begin{figure}
\includegraphics[bb=0 0 612 500,scale=0.28]{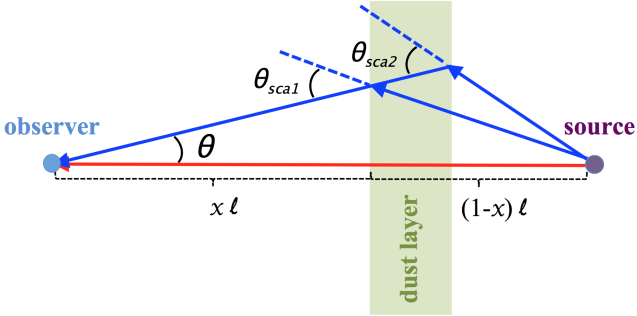} 
\caption{Geometry of the dust scattering by a single thick dust layer. $\theta$ is the viewing angle. $\theta_{sca1}$ and $\theta_{sca2}$ are the two scattering angles at the same viewing angle but different location in the dust layer. $\ell$ is the absolute distance between the source and the observer, $x$ is the fractional distance from the dust layer to the observer.}
\label{fig-cartoon1}
\end{figure}

\subsubsection{Basic Principles of the Dust Scattering Theory}
We assume a point source that is located at a distance $\ell$ from the observer characterized by a flux $F_{X}$ and absorbed by neutral material with a hydrogen column density $N_{H}$. With assumptions made in the dust grain models about the dust-to-gas ratio and abundances, this $N_{H}$ is determined by the total number of dust grains required (hereafter: $N_{H,sca}$). Moreover, we assume that dust grains are distributed along the LOS with a normalized distribution function $f(x)$ (where $x$ ($0\le x \le1$) is the fractional distance between the observer and the dust grains, as drawn in Fig.\ref{fig-cartoon1}). The intensity of the scattering light at the viewing angle $\theta$, due to a single dust scattering (Mathis \& Lee 1991), is given by:
\begin{multline}
\label{equ-radp}
I^{(1)}_{sca}(\theta)=F_{X} N_{H,sca} {\int^{E_{max}}_{E_{min}}} S(E) {\int^1_0} {\frac{f(x)}{(1-x)^2}} {\int^{a_{max}}_{a_{min}}} n(a) \\
\times{\frac{d\sigma_{sca}(a,x,E,\theta)}{d\Omega}}~da~dx~dE
\end{multline}
where $S(E)$, $n(a)$, $\sigma_{sca}(a,x,E,\theta)$ are the normalized source spectrum, dust grain size distribution and the scattering cross section, respectively. $n(a)$ depends on the dust grain model and contains assumptions about the abundances and the dust-to-gas ratio. The wavelength of the X-ray photon is much smaller than the typical size range of dust grains in the ISM, and the refraction index of the dust grain is close to unity. Therefore, the Rayleigh-Gans approximation, instead of a full Mie calculation, can be used to simplify the calculation (e.g. Mathis \& Lee 1991). So the differential cross section can be written as:
\begin{multline}
\label{equ-difcs}
\frac{d\sigma_{sca}}{d\Omega} = (1.1~cm^2sr^{-1})\Big{(}\frac{2Z}{M}\Big{)}^2\Big{(}\frac{\rho}{3~g~cm^{-3}}\Big{)}\Big{[}\frac{F(E)}{Z}\Big{]}^2\\
\times\Big{(}\frac{a}{1~\mu m}\Big{)}^6~\Phi^2(a,x,E,\theta))
\end{multline}
where $\Phi(a,x,E,\theta)$ is the form factor (Bohren \& Huffman 1983) that can be approximated by a Gaussian form for a spherical dust grain with both good accuracy and high computing efficiency:
\begin{multline}
\label{equ-theta1}
\Phi^2(a,x,E,\theta)=\\
exp~[-0.4575~E(keV)^2~a(\mu m)^2\theta_{sca}(arcmin)^2]
\end{multline}
where the scattering angle $\theta_{sca}$ can be derived from $x$ and $\theta$ from simple geometric calculations:
\begin{equation}
\label{equ-theta2}
\theta_{sca}=arctan\Big{(}\frac{x}{1-x} \cdot tan\theta\Big{)} + \theta
\end{equation}
The advantage of using the Gaussian approximation is the much higher computing efficiency with good accuracy. Mathis \& Lee (1991) compared the scattering profiles between the exact form factor and the Gaussian approximation, and found at most 10 percent difference at all scattering angles. According to the above equations, we can also define a typical dust grain size for a specific $\theta_{sca}$ as:
\begin{equation}
\label{equ-awing}
a(\mu m)=\frac{1.4785}{E(keV)~\theta_{sca}(arcmin)}
\end{equation}
Note that for a dust layer with a non-negligible thickness, the scattering light at a specific viewing angle $\theta$ comprises a range of scattering angles ($\theta_{sca1}-\theta_{sca2}$, Fig.\ref{fig-cartoon1}). It can therefore be predicted that the dust scattering from a distinct layer will produce a halo around X-ray sources. Eq.\ref{equ-theta1} indicates that the halo intensity is the strongest in the core and fades out quickly outside a few arcmin, making a point source slightly extended. The typical halo size changes with the photon energy as $\propto exp({-E^{2}}$). Therefore, a strong dust scattering halo is most likely to be observed in the soft X-ray band from a bright X-ray source with a significant amount of intervening dust on the LOS.

Finally, the optical depth of the dust scattering $\tau_{sca}$ can be estimated from the observation using the fraction of the halo flux ($F_{sca}$) in the total observed flux ($F_{obs}$) (Mathis \& Lee 1991; Predehl \& Schmitt 1995; Valencic et al. 2009):
\begin{equation}
\label{equ-tau}
\tau_{sca} = - ln~(1- F_{sca}/F_{obs})
\end{equation}

\subsubsection{Variability of the Dust Scattering Halo}
\label{sec-eclipse}
A dust scattering halo will show variability if the source is variable (Tr\"{u}mper \& Sch\"{o}nfelder 1973; Mao, Ling \& Zhang 2014; Heinz et al. 2015; Vasilopoulos \& Petropoulou 2016). Time lags are caused by the arrival time differences of scattering photons with different light paths (Xu, McCray \& Kelley 1986). Based on simple geometrical calculations, the time-lag of a scattering photon relative to an un-scattered photon can be derived as:
\begin{equation}
\label{equ-delta1}
\Delta t(x,\theta,\ell)=\frac{\ell}{c}~\Big{(}\sqrt{\Big{(}\frac{x}{cos\theta}\Big{)}^2-2x+1}+\frac{x}{cos\theta}-1\Big{)}
\end{equation}
where $c$ is the speed of light. Since dust scattering mainly happens at small angles, ${\Delta}t(x,\theta,\ell)$ can be simplified with small angle approximations:
\begin{equation}
\label{equ-delta2}
\Delta t(x,\theta,\ell)\approx(1.21s)~\frac{x}{1-x}~\theta(arcsec)^2~\ell(kpc)
\end{equation}
According to these equations, for a viewing angle of 2 arcsec to a dust grain at $x=0.9$ from a source at $\ell=8~kpc$, the lag is 350 s. These equations also show that time-lag increases with increasing viewing angles. For a fixed viewing angle, the time-lag increases as the scattering dust is located closer to the source. The lag scales linearly with the absolute source distance.

\begin{table*}
 \centering
  \begin{minipage}{175mm}
  \centering
   \caption{List of \xmm\ and \cxo\ observations used for this paper. The list of observations during which \axj\ was in quiescence can be found in the Appendix~\ref{app-obslis}. `Exp' is the total exposure time and `CL-Exp' is the good exposure time after subtracting the periods of background flares. $\theta_{off}$ is the off-axis angle of \axj. $r_{pileup}$ is the inner circular radius of the PSF affected by photon pileup (Section~\ref{sec-data}). `PFW': Primary Full Window mode. `HETG': High Energy Transmission Grating.}
    \label{tab-obs}
     \begin{tabular}{@{}ccccccccc@{}}
     \hline
     \cxo\ &&&&&&&&\\
     Obs-ID & Obs-Date  & Target &ACIS (Grating) & Exp & CL-Exp & $\theta_{off}$ & $r_{pileup}$\\
     & & & & (ks) & (ks) & (arcmin) & (arcsec)\\
     \hline
     9169& 2008-05-05 &\sgra\ &ACIS-I (None)&27.6&27.6&1.80 &2.0\\
     9170&2008-05-06 &\sgra\ &ACIS-I (None)&26.8&26.8&1.80 &2.0\\
     9171&2008-05-10 &\sgra\ &ACIS-I (None)&27.7&27.7&1.80 &2.1\\
     9172&2008-05-11 &\sgra\ &ACIS-I (None)&27.4&27.4&1.80 &2.2\\
     9174&2008-07-25 &\sgra\ &ACIS-I (None)&28.8&28.8&1.32 &2.0\\
     9173&2008-07-26 &\sgra\ &ACIS-I (None)&27.8&27.8&1.32 &2.0\\
     17857&2015-08-11 &AX J1745-2901 & ACIS-S (HETG)&117.2&111.1&0.34&2.4\\
     \hline
     \xmm\ &&&&&&&&\\
     Obs-ID & Obs-Date  &  Target & pn-Filter & Exp & CL-Exp & $\theta_{off}$ & $r_{pileup}$\\
     & & & & (ks) & (ks) & (arcmin)  & (arcsec)\\
     \hline
     0402430701 &2007-03-30 &\sgra\ &PFW-Medium&34.2&21.3&2.57 &20\\
     0402430301 &2007-04-01 &\sgra\ &PFW-Medium&105.4&56.1&2.55 &20\\
     0402430401 &2007-04-03 &\sgra\ &PFW-Medium&105.8&38.1&2.60 &20\\
     0505670101 &2008-03-23 &\sgra\ &PFW-Medium&105.7&64.5&2.57 &20\\
     0724210201 &2013-08-30 &\sgra\ &PFW-Medium&58.5&53.4&1.94 &20\\
     0700980101 &2013-09-10 &\sgra\ &PFW-Medium&38.7&35.8&1.92 &20\\
     0724210501 &2013-09-22 &\sgra\ &PFW-Medium&43.9&32.2&1.89 &20\\
     0743630801 &2015-04-01 &\sgra\ &PFW-Medium&27.0&23.4&2.56 &20\\
     \hline
     \end{tabular}
 \end{minipage}
\end{table*}

\subsection{Dust Grain Models}
\label{sec-grainmodel}
The modelling of the dust scattering halo strongly depends on the dust grain model (e.g. Xiang et al. 2005; Valencic \& Smith 2015). The classic dust grain model, proposed by Mathis, Rumpl \& Nordsieck (1977) (hereafter: MRN77 dust grain), assumes a power law form for the size distribution of graphite and silicate grains with a size range of 0.005-0.25$\mu$m. Later a smaller-size carbonaceous grain population (in the form of Polycyclic Aromatic Hydrocarbon - PAH) was added to explain the infrared (IR) emission features and to reproduce the 2175\AA\ hump in the ultraviolet (UV) extinction curve (Li \& Draine 2001; Weingartner \& Draine 2001 - WD01 dust grains for different $R_{V}$ (def. $A(V)/E(B-V)$) and carbon abundance; Draine 2003). Zubko, Dwek \& Arendt (2004) (hereafter: ZDA04) considered various forms of carbon in the ISM, different abundances, and composite particles encompassing organic refractory material (C$_{25}$H$_{25}$O$_{5}$N), water ice (H$_{2}$O) and voids, and proposed 15 dust grain models which can all reproduce the $R_{V}=3.1$ extinction curve. Moreover, Xiang et al. (2011) proposed a different dust grain model (XLNW) where the olivine grains in the BARE-GR-S dust grain model (BARE: without composite grains; GR: graphite; S: solar abundances) of ZDA04 is replaced with a new type of grain comprising a metallic iron core and a troilite/enstatite shell.

Due to the lack of knowledge about the dust grain composition and size distribution in different locations of the ISM, previous studies about the dust scattering halo often adopted various dust grain models and compared the results. It was reported that the BARE- dust grains in ZDA04 were better than COMP- dust grains (with composite grains) in terms of fitting the halo around sources whose LOS has a short overlap with the Galactic plane (Valencic et al. 2009;  Xiang et al.  2011; Valencic \& Smith 2015). For the GC LOS, Fritz et al. (2011) measured the extinction curve within 1-19 ${\mu}m$ and fitted various dust grain models to the IR extinction curve, and reported the necessity of adding composite dust grains containing H$_{2}$O to the pure carbonaceous and silicate grains. Therefore, the COMP-AC-S dust grain model (AC: amorphous carbon) of ZDA04 was proposed, for the first time, to be the best dust grain model for the GC LOS.

\subsection{\axj}
\label{sec-axj}
The GC region is filled with intense and complex diffuse emission, so a bright X-ray source is required to produce a detectable bright dust scattering halo that overwhelms the diffuse emission. Then the intrinsic halo profile can be extracted with enough signal-to-noise (S/N) to study the dust properties along the LOS. \axj\ is an ideal source because it can be very bright in X-rays. It is a low mass X-ray binary (LMXB) consisting of a neutron star and a K0V type companion star (Ponti et al. 2015b). It was first discovered during {\it ASCA} observations of the GC in 1993 as an eclipsing transient (Kennea \& Skinner 1996; Maeda et al. 1996). \cxo\ observations provided precise measurements of its coordinate (17:45:35.64, -29:01:33.90 with a 1 $\sigma$ coordinate error of 0.32\arcsec)\footnote{Degenaar \& Wijnands (2009) report a similar \cxo\ position with a 1 $\sigma$ coordinate error of 0.6\arcsec. To improve the astrometric accuracy, we choose a long \cxo\ observation (ObsID:03665, 89.9 ks exposure time) where \axj\ is faint (so it is free from photon pile-up) but still have 285 counts for an accurate position measurement. We find three bright stars in the Tycho-2 catalog detected in the soft X-ray band within the \cxo\ field-of-view, and use them to perform a high-precision astrometry correction, which then allows us to reduce the coordinate error by a factor of 2.}, which is only 1.45 arcmin from \sgra\ (Heinke et al. 2009). Therefore, it has been intensely observed by many X-ray observatories such as \cxo, \xmm, {\it Swift} and {\it Suzaku} during GC monitoring campaigns (Hyodo et al. 2009; Degenaar et al. 2012; Ponti et al. 2015a). 

Previous observations revealed periodic eclipsing and irregular dipping in the light curve of \axj, indicating a high inclination angle of this binary system (Maeda et al. 1996) with an orbital period of 30063.74$\pm$0.14 s and an eclipse duration of $\sim$1440 s (Hyodo et al. 2009). Analysing more than 20 years of \xmm\ and {\it ASCA} observations, Ponti et al. (2017a) measured a long-term decrease of the orbital period of $(4.03\pm0.32)\times10^{-12} s/s$ as well as significant jitter.

\axj\ is one of the brightest X-ray transient within a few arcmin of \sgra (Degenaar et al. 2012; Ponti et al. 2015b). Its luminosity was observed to reach a few percent Eddington luminosity assuming that the source is inside the GC region (i.e. within a few hundred pc from Sgr A$^\star$), which is within the typical luminosity range for `atoll' sources (Gladstone, Done \& Gierli\'{n}ski 2007). The $N_{H,abs}$ towards \axj\ is measured to be $\sim2.0\times10^{23}cm^{-2}$ by Ponti et al. (2015b) from fitting the \xmm\ spectra with the {\sc Xspec} {\sc phabs} absorption model and solar abundances of Anders \& Grevesse (1989) (hereafter: AG89). But Paizis et al. (2015) fitted the \cxo\ spectra with the {\sc tbabs} model and Interstellar Medium (ISM) abundances of Wilms, Allen \& McCray (2000) (hereafter: WAM00), and reported $N_{H,abs}=3.4\times10^{23}cm^{-2}$. The difference mainly arise from different assumptions about the abundances. In comparison, the $N_{H,abs}$ measured from \sgra's X-ray flares is $1.23\times10^{23}cm^{-2}$ using the {\sc wabs} model and Anders \& Ebihara (1982) solar abundances (Porquet et al. 2008), or $1.5\times10^{23}cm^{-2}$ using the {\sc tbnew} model and WAM00 ISM abundances (Nowak et al. 2012), which is also consistent with $1.6\times10^{23}cm^{-2}$ measured most recently by Ponti et al. (2017b) using the same absorption model and abundances. These $N_{H,abs}$ measurements are all significantly smaller than that of \axj. Moreover, Degenaar \&  Wijnands (2009) reported the brightest X-ray burst from this source reaching a luminosity of $1.3\times10^{38}erg~s^{-1}$ for a GC distance of 8 kpc, approaching the Eddington luminosity of a typical neutron star (e.g. Kuulkers et al. 2003). Therefore, it can be inferred that \axj\ is likely located inside or behind the GC.


\begin{figure*}
\begin{tabular}{ccc}
\includegraphics[bb=0 0 612 660, scale=0.25]{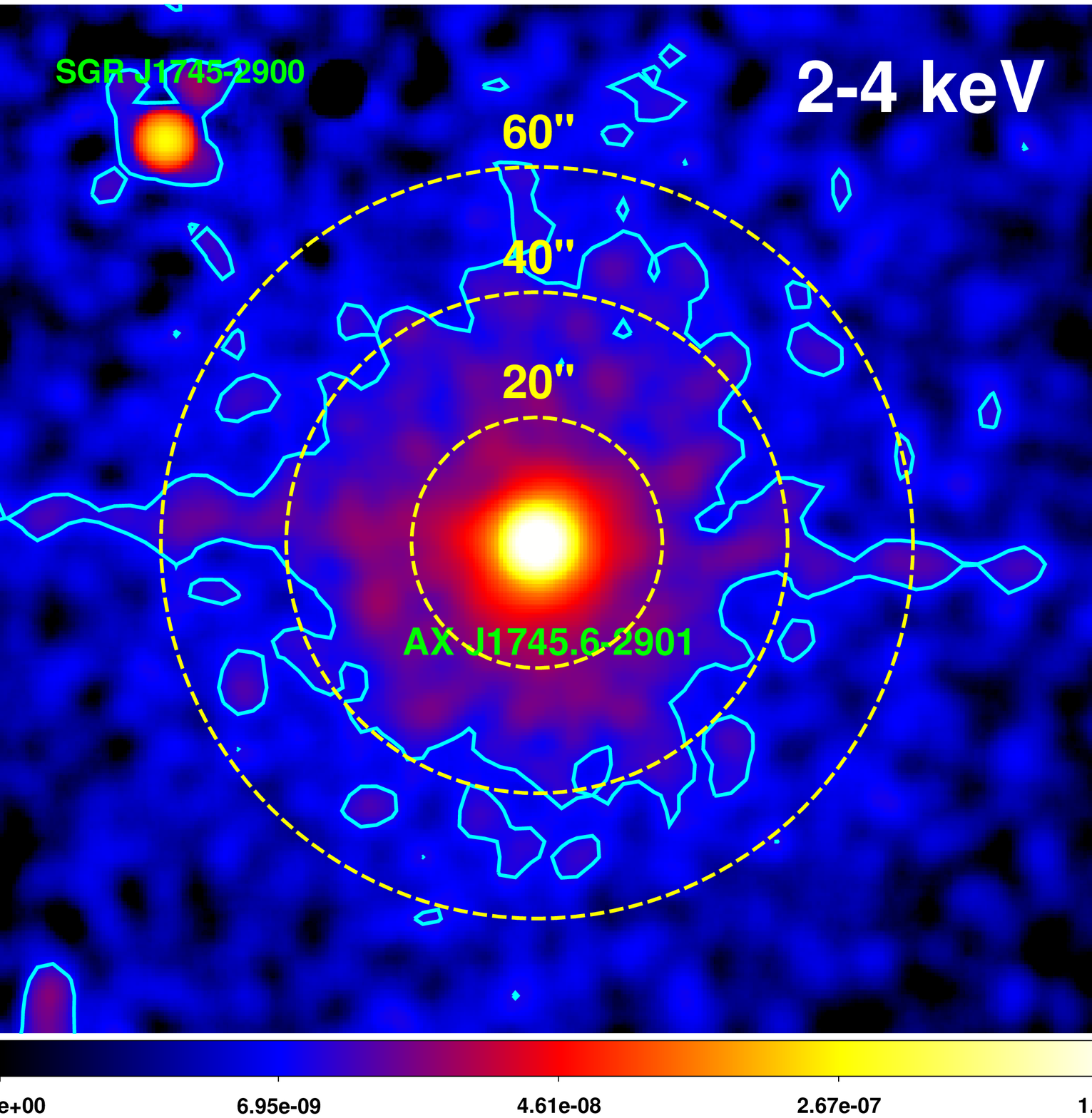} &
\includegraphics[bb=0 0 612 660, scale=0.25]{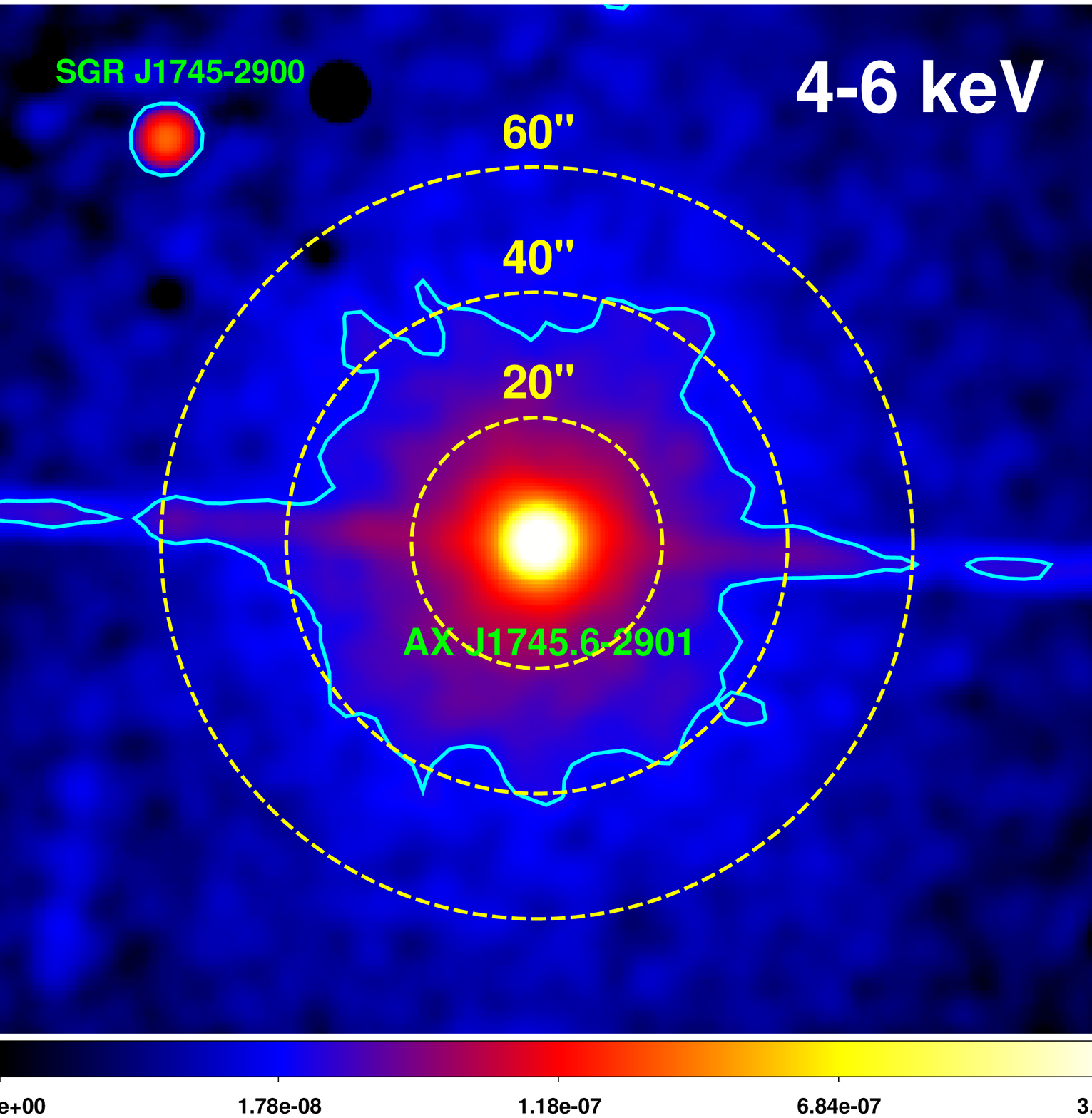} &
\includegraphics[bb=0 0 612 660, scale=0.25]{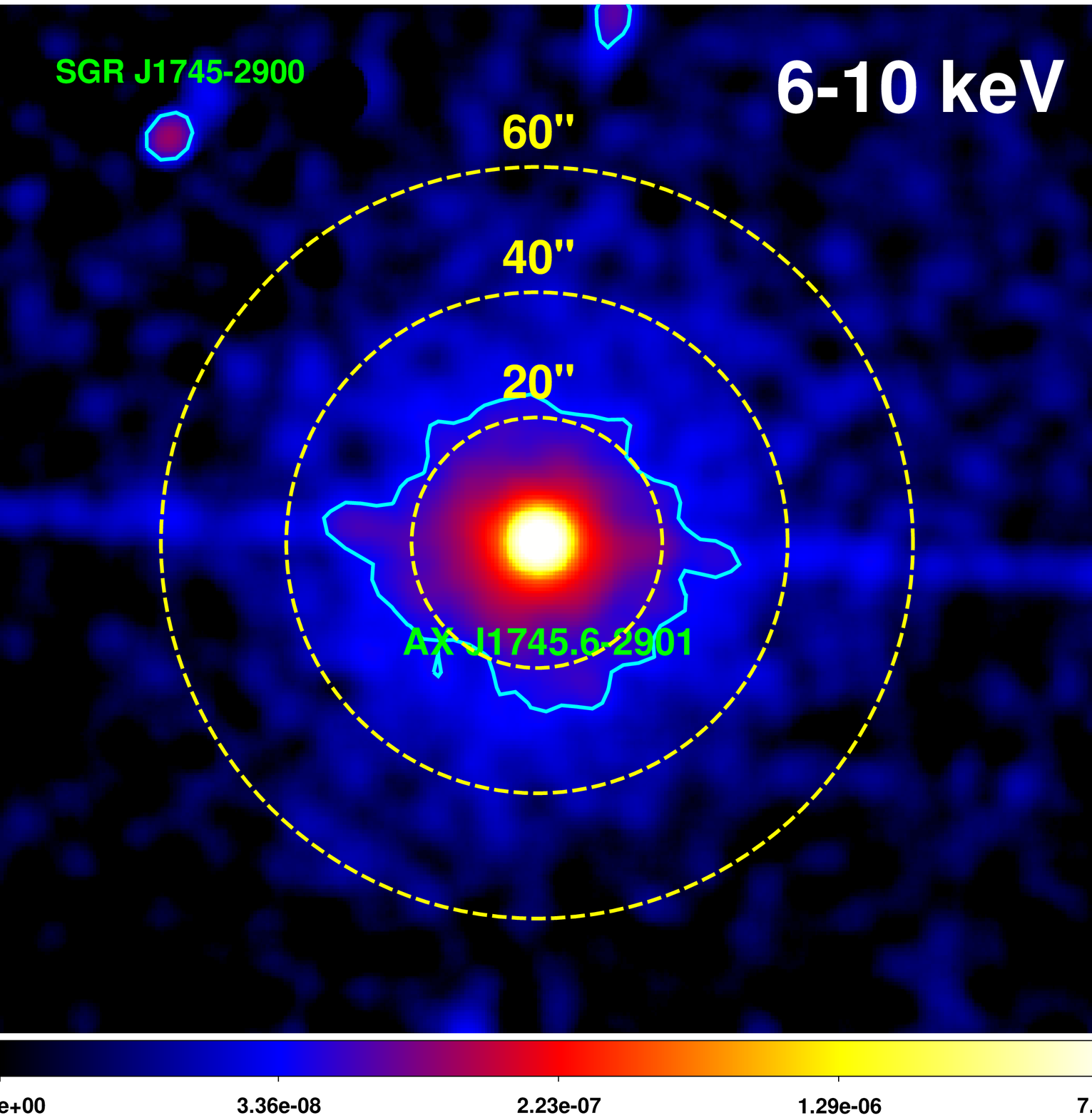} \\
\end{tabular}
\caption{{\it Chandra} ACIS-S (ObsID:17857) flux images of \axj\ in the 2-4, 4-6 and 6-10 keV bands, showing the shrink of the dust scattering halo towards higher energies. The background based on 12 ACIS-S observations with \axj\ in quiescentcehas been subtracted (see Section~\ref{sec-data}). In every panel, the cyan contour indicates the location where the surface brightness is only 0.4\% of the value at 2.5 arcsec from the core (inside which more than 1\% pile-up is expected). The dashed yellow circles indicate regions with 20, 40 and 60 arcsec radius centred on \axj. Note that the horizontal structure across \axj\ is due to the readout streak.}
\label{fig-halo}
\end{figure*}

This paper presents a detailed analysis of the dust scattering halo around \axj. We organize the paper as follows. In Section 2 we describe the \cxo\ and \xmm\ observations of \axj\ and
data reduction, as well as the steps to extract the dust scattering halo around \axj.
Section 3 describes the point spread function (PSF) of \cxo\ and \xmm.
Section 4 presents a detailed modelling of the dust scattering halo
in order to explore the properties of the dust grains along the LOS. The effect of the
dust scattering on the observed spectra is studied in Section 5.
Discussions about the dust properties towards \axj\ and more details about the halo profile are all included in Section 6.
The final section summarizes our work and provides the main
conclusions. Throughout this paper, we adopt a distance of 8 kpc to the GC (Reid et al. 2009; Genzel et al. 2010; 	
Boehle et al. 2016), the uncertainty of which should not affect the results of this work.

\section{Observation and Data Reduction}
\label{sec-data}
\axj\ is one of the brightest transients within a few arcmin from Sgr A$^\star$. Being so close, it is in the field of view (FoV) of many observations targeting Sgr A$^\star$. In this work we used archival and new \cxo\ and \xmm\ observations. \cxo\ has high spatial resolution and small PSF, which can produce high resolution halo radial profiles; while \xmm\ provides a large effective area and FoV, which leads to a significant detection of the dust scattering halo up to 10 arcmin. Therefore, combining the data from these two satellites is ideal to study the halo profile.

\subsection{\cxo\ Observations}
\label{sec-obs-cxo}
The \cxo\ data website was used to search for observations of \axj\footnote{http://cda.harvard.edu/chaser}. We selected observations where \axj\ was observed at $\le3$ arcmin off-axis angle\footnote{the angle between the HRMA optical axis and the source position} because of significant PSF degradation at larger off-axis angles. Among these observations, we chose those where \axj\ was bright enough to show a clear halo above 2 keV ($L_{2-4keV}\ge10^{-12}~erg~cm^{-2}~s^{-1}$). Then we dropped observations with short exposures ($\lesssim5~ks$), as these observations were too short to constrain the halo profile. Our selection resulted in 6 ACIS-I observations targeting \sgra\ and 1 recent ACIS-S observation targeting \axj\ (Table~\ref{tab-obs}).

In order to measure the detector background and diffuse emission underneath \axj\, we made use of observations where \axj\ was at $\le3$ arcmin off-axis angle but was in the quiescent state. To ensure that the non-detection of \axj\ was due to its intrinsic weakness rather than the short exposure time, we chose observations with more than 15 ks exposure and found 17 such observations in the ACIS-I mode (with 713.3 ks clean exposure in total) and 12 observations in the ACIS-S mode (with 702.7 ks clean exposure in total) (Table~\ref{app-tab-obs}). The long accumulated exposure time allows an accurate determination of both the detector background and the diffuse emission in this region. Moreover, we checked that \axj\ was located in the same type of CCD chip, i.e. Front Illuminated (FI) chip in ACIS-I and Back Illuminated (BI) chip S3 in ACIS-S. This is to ensure the consistency of the detector background among all observations in ACIS-I and ACIS-S, separately.

\begin{figure*}
\begin{tabular}{cc}
\includegraphics[scale=0.48]{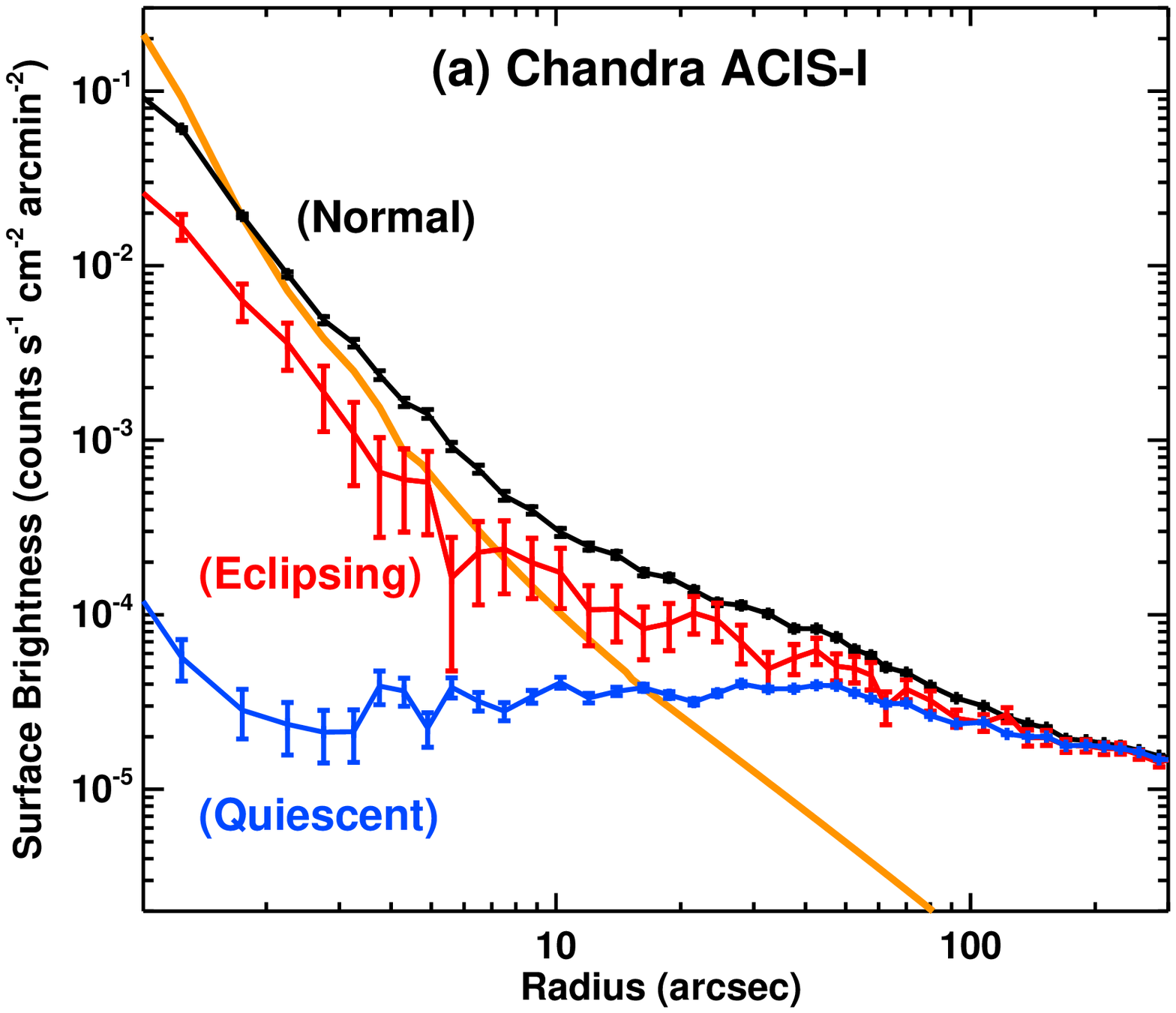} &
\includegraphics[scale=0.48]{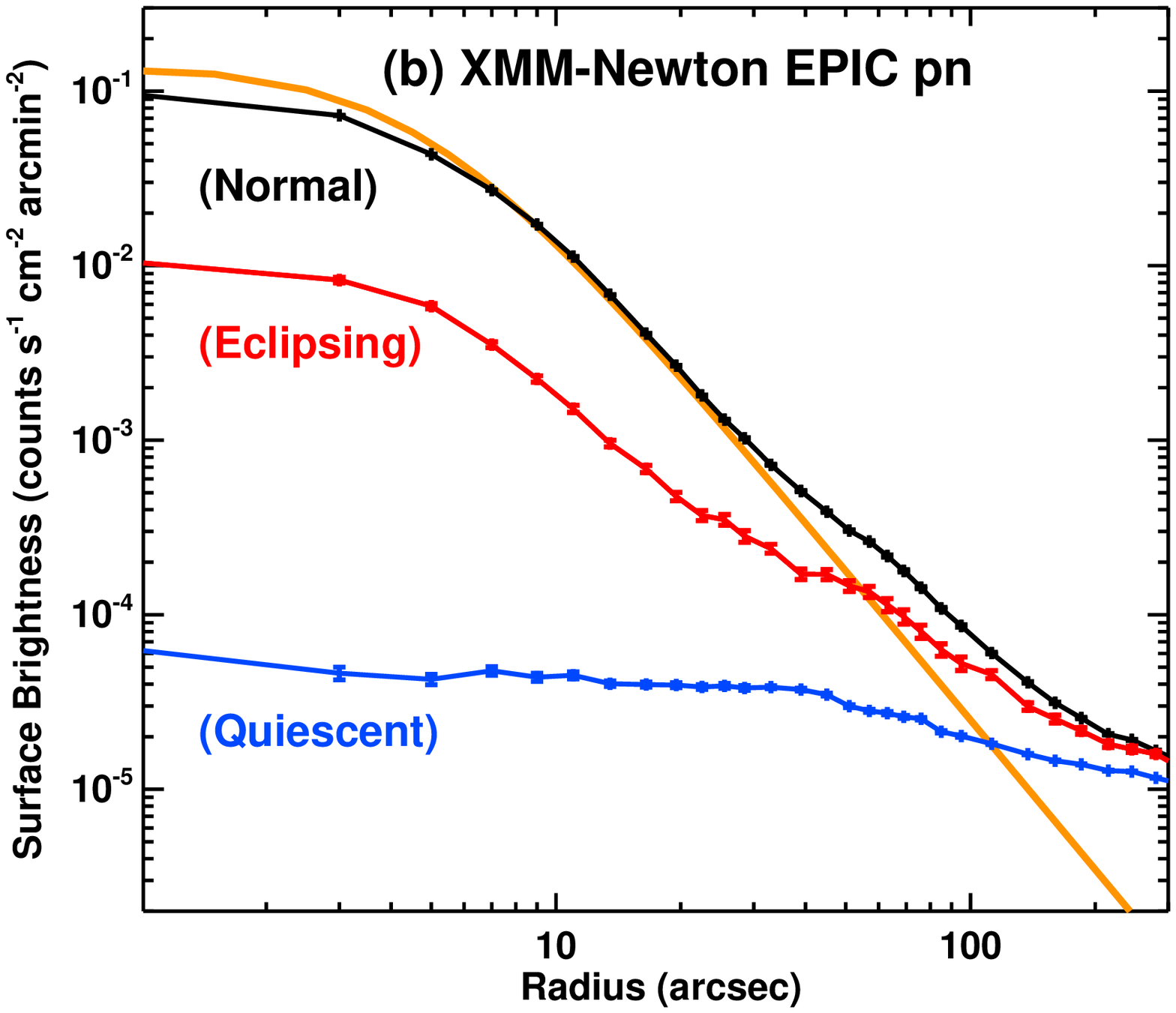} \\
\end{tabular}
\caption{Radial profile of \axj\ in the normal (black) and eclipsing phases (red) in the 4-6 keV band. Panel-a shows the combined radial profile from 6 \cxo\ ACIS-I observations in Table~\ref{tab-obs}. The quiescent profile (blue) is based on a combination of 17 Chandra ACIS-I observations where \axj\ was in quiescent. The slight rising of the radial profile within 2 arcsec is due to the weak emission from \axj\ when several observations were combined. Panel-b shows the combined radial profile from 8 \xmm\ observations in Table~\ref{tab-obs}. The emission inside the eclipsing phase comes from the dust scattering. In both panels, the solid orange profile shows the instrumental PSF at 5 keV. The difference between the halo radial profiles of \cxo\ and \xmm\ is caused by their different PSF. }
\label{fig-radprof}
\end{figure*}

The \cxo\ software {\sc CIAO} (v4.8.1) and Chandra Calibration Database (CALDB, v4.7.2) were used to analyze all the \cxo\ observations following the standard data reduction steps (see e.g. Clavel et al. 2013 for detailed descriptions). Firstly, the {\sc chandra\_repro} script was used to reprocess all the ACIS data. To ensure the astrometry consistency among all the observations, we followed the standard thread to perform the astrometry correction by taking ObsID:14468 (146 ks exposure time) as the reference observation. For every observation, we used the {\sc wavdetect} script to perform the point source detection and all the point sources were excluded before extracting the background light curve, which was then used as the input for the {\sc deflare} script to remove the background flares from the event file. \axj\ was piled up in all the selected observations, so we used the {\sc pileup\_map} script in {\sc CIAO} (v4.8.1) to estimate the size of the 1\% pile-up region around \axj\ (Table~\ref{tab-obs}). All data within this pile-up radius were removed from our study. Furthermore, we excluded the data when \axj\ was in the eclipsing phase (detected in 5 observations, ObsID: 9169, 9170, 9173, 9174, 17857) and 300 s after the eclipse egress time. This is because the halo profile was varying due to a delayed response to the eclipsed signal which needs to be studied independently (Jin et al. in prep.).

We used the {\sc merge\_obs} script and {\sc reproject\_image\_grid} script in the \cxo\ software to create images and exposure maps (including the vignetting effect), re-project the images and combine images from all observations. The {\sc dmimgcalc} script was used to perform background subtraction. Fig.\ref{fig-halo} shows the final flux images of \axj\ in 2-4, 4-6 and 6-10 keV bands from the ACIS-S observation (ObsID: 17857). It is clear that \axj\ appears much more extended than a point-like source, and the extension has a clear energy-dependence, consistent with the prediction of the dust scattering theory.

To extract the radial profile of \axj, we first excluded all other point sources including transients in every observations, then an elliptical region of 2\arcmin$\times$6\arcmin\ was used to exclude the region around \sgra\ where the diffuse emission is strongest (e.g. Heard \& Warwick 2013a,b; Ponti et al. 2015a). All artificial features such as the readout streak and the HETG arms (in ACIS-S observations) were also masked out by employing box regions of 7 arcsec wide from the edge of the FoV down to 2 arcsec from \axj. We defined a set of annulus regions, and used the {\sc Funtools} in {\it ds9} to extract the photon counts and exposure in every annulus region from the image and exposure map of every observation\footnote{also see http://cxc.harvard.edu/ciao/threads/radial\_profile}. The combined radial profile was obtained by adding counts from all observations and dividing it by the total exposure time in every annulus region. The Poisson counting errors are small because of the accumulated number of counts from all the observations, but there are additional small dispersions in the shape of radial profiles from different observations, which are likely due to the weak dependence of PSF on the small variation of off-axis angles (Table~\ref{tab-obs}), and/or the fluctuation of residual background contamination from e.g. the extended PSF wing of bright sources and diffuse emission in the FoV. Therefore, we used the MPFITEXY routine (Williams, Bureau \& Cappellari 2010, Markwardt 2009) to calculate error bars for the combined radial profile, which takes into account both the Poisson error and the small dispersion between different observations (see Tremaine et al. 2002 for the statistical method).

The above procedure was repeated to create the radial profile of \axj\ as well as the radial profile of the underlying diffuse emission and detector background when \axj\ was in quiescence (Fig.\ref{fig-radprof}a, also see Fig.\ref{app-fig-radprof} for the energy-dependence of the radial profile). The intrinsic radial profile of \axj\ was derived by subtracting the combined radial profile observed when \axj\ was in quiescence from the one when \axj\ was bright, with the standard error propagations. Note that we did not combine the radial profiles from ACIS-I and ACIS-S observations, because the ACIS-S observation was in the sub-array mode. \axj\ was weakly detected in the combined background radial profile within 2 arcsec due to the enhanced S/N. We also repeated the above procedures to extract the mean radial profile of \axj\ during its eclipsing phase (Fig.\ref{fig-radprof}a red points), which allows us to observe the shape of the dust scattering halo without the PSF of \axj\ itself. However, the light curve of \axj\ in Ponti et al. (2015) shows that the halo is variable during the eclipsing phase due to the time-lag effect (see Section~\ref{sec-eclipse}), so in this work we do not model the halo profile in the eclipsing phase.

\subsection{\xmm\ Observations}
\label{sec-obs-xmm}
We used the \xmm\ Science Archive (XSA) to search for \xmm\ (Jansen et al. 2001) observations of \axj\ with similar selection criteria as for the \cxo\ observations. We found 9 observations where \axj\ was observed at $\le3$ arcmin off-axis angle where the PSF was not degraded significantly, and the source was bright enough for a significant halo detection above 2 keV ($L_{2-4keV}\ge10^{-11}~erg~cm^{-2}~s^{-1}$). All these observations were in the same mode and targeted \sgra\ so that \axj\ was at the same off-axis angle. In order to determine the underlying diffuse emission and detector background, we selected another 14 observations targeting \sgra\ with $\gtrsim$15 ks exposure time but with no detection of \axj\ (so it was in quiescence). The accumulated exposure time is 358.3 ks, which can ensure an accurate measurement of the diffuse emission and detector background.

For each of these observations, we used the \xmm\ Science Analysis System ({\sc SAS} v15.0.0) to process the data and applied the most recent calibrations. We only focused on the EPIC-pn data as pn provides the highest count rate among all three EPIC cameras. The {\sc epchain} task was used to reprocess the data. Then we excluded all high background periods detected from the 7-15 keV background light curve, and followed the method in Urban et al. (2011) (also see Leccardi \& Molendi 2008) to check and ensure no significant soft proton residuals. We excluded one observation of \axj\ (ObsID: 0504940201), as the clean exposure time was only 6 ks. The remaining 8 observations of \axj\ are listed in Table~\ref{tab-obs}. The 14 observations that caught \axj\ in quiescence are listed in Table~\ref{app-tab-obs}, with a total of 332.1 ks clean exposure time.

\begin{figure}
\centering
\includegraphics[scale=0.45]{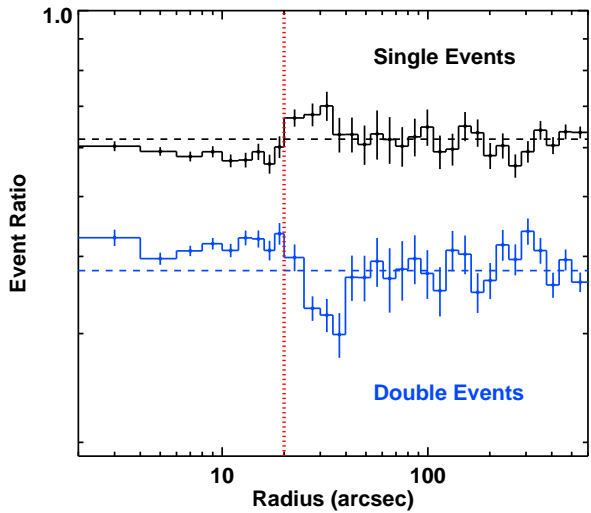}
\caption{Radius dependence of the ratio between single-events and all events (black), and between double-events and all events (blue) for 7-10 keV in \xmm\ (OBSID: 0743630801). The two horizontal dashed lines indicate the expectation of no pile-up effect. The vertical red dotted line indicates 20 arcsec radius within which the pile-up effect is seen. }
\label{fig-xmmpileup}
\end{figure}

We used the {\sc epatplot} task to determine the pile-up region, and excluded data within 20 arcsec of \axj\ to safely avoid the pile-up effect. We double-checked this radius by plotting the radial dependence the ratios of single-events and double events in all events for the observation (OBSID: 0743630801) where \axj\ had the highest flux (see Costantini, Freyberg \& Predehl 2005). For the observed spectrum of \axj, the pile-up effect is most obvious above 7 keV, and so we used events within 7-10 keV to perform this test. Fig.\ref{fig-xmmpileup} shows that, within 20 arcsec, the pile-up effect causes single-event's ratio to be lower than expectation, and double-event's ratio to be higher than expectation.

For every observation, the 2-10 keV light curve of \axj\ was extracted with the {\sc evselect} task. The light curve was used to search for eclipsing periods and the latter were excluded from the event file. We also carefully excluded some dipping periods. Then we used the Extended Source Analysis Software ({\sc ESAS}; Snowden \& Kuntz 2011) in {\sc SAS} (v15.0.0) to create and process images from the event file. We performed point source detection using the {\sc cheese} task. Then the {\sc pn\_spectra} task was used to create the normal and Out-of-Time (OoT) processing images and exposure maps (including the vignetting effect). The OoT image was multiplied by 0.063 and subtracted from the normal image using the {\sc farith} task ({\sc FTOOLS} v6.19). The {\sc pn\_back} task was used to create the particle background image and then subtracted this from the normal image.

All point sources except \axj\ were excluded in every observation before extracting the radial profile. An elliptical region of 12\arcmin$\times$16\arcmin\ was used to exclude the strong diffuse emission around \sgra. We note that there were still residual OoT features in the final image of some observations, which was because \axj\ was piled up so that the correction was not accurate. Therefore, we visually checked all images and, when necessary, we used a rectangular region to mask out the OoT residual from the CCD edge down to 20 arcsec from \axj. We chose the width of the rectangle up to 250 arcsec depending on the strength of the OoT residual feature. Some weak features near the edge of FoV due to stray light were also visually masked out. The exposure map created by the {\sc pn\_back} task was not multiplied by the effective area, so we used the {\sc SAS} task {\sc psfgen} to produce the unvignetted Ancillary Response File (ARF), from which we obtained the spectral shape weighted effective area in every energy band. Finally, we obtain the combined radial profiles from all the observations and performed background subtraction using the same method as for \cxo\ (see Section~\ref{sec-obs-cxo} and Fig.\ref{fig-radprof}b).

\begin{figure}
\centering
\includegraphics[scale=0.48]{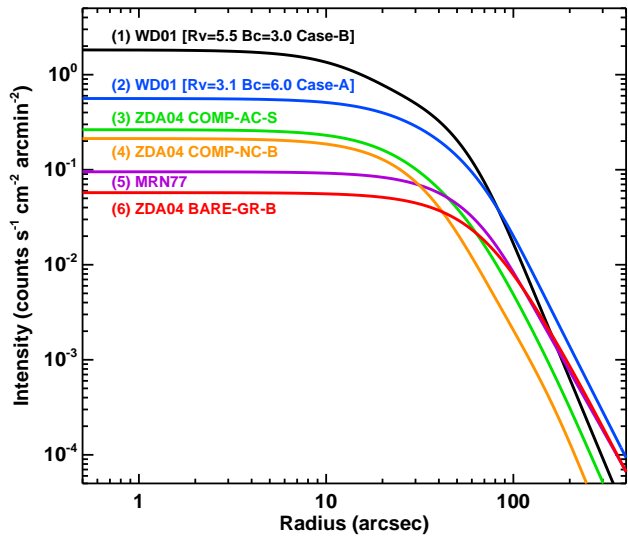}
\caption{Radial profile comparison for some representative dust grain models. These radial profile models are for a source at 8.15 kpc with a flux of unity and $E~=~5~keV$. The light passes through a single smoothly distributed dust layer between 2.5 and 3.5 kpc with a total $N_{H,sca}~=~1.5\times10^{23}~cm^{-2}$. No instrumental PSF was convolved. These radial profiles reveal the typical halo shape diversity among the dust grain models in MRN77, ZDA04 and WD01.}
\label{fig-haloshape}
\end{figure}

\section{PSF of \cxo\ and \xmm}
The image of \axj\ consists of a central point source plus a dust scattering halo, convolved with a 2-dimensional (2D) instrumental PSF. This 2D convolution can significantly smooth the halo profile especially when the PSF is broad such as in \xmm.

\begin{figure}
\centering
\includegraphics[bb=40 120 540 648, scale=0.43]{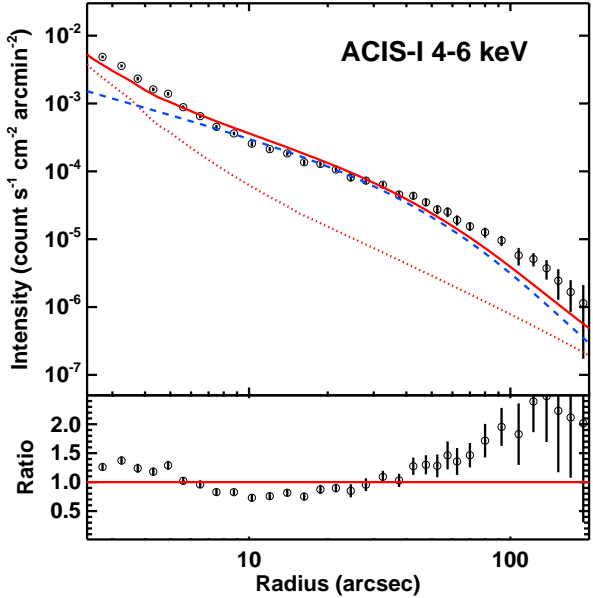} 
\caption{Radial profile observed by \cxo\ ACIS-I in the 4-6 keV band. The model comprises the ACIS-I PSF (red dotted curve) and a single dust layer smoothly distributed from Earth to \axj (blue dashed curve), which used the COMP-AC-S dust grain model. The lower panel shows the residuals.}
\label{fig-1dl}
\end{figure}

\subsection{PSF of \cxo\ ACIS}
The on-axis half energy width of the PSF in the \cxo\ High-Resolution Mirror Assembly (HRMA)/Advanced CCD Imaging Spectrometer (ACIS) is 0.4-0.7 arcsec within 0.3-10 keV (\cxo\ Proposers' Observatory Guide v18.0\footnote{http://cxc.harvard.edu/proposer/POG/html/chap6.html\#tth\_sEc6.6}), indicating that the halo profile detected by \cxo\ ACIS is heavily smeared by the PSF convolution (Smith, Edgar \& Shafer 2002; Smith 2008). Therefore, an accurate PSF is mainly required to disentangle the point source from the dust scattering halo. We used the web-based {\sc ChaRT} software to perform the ray-trace simulation which takes parameters directly from real observations (ObsID:09174 for ACIS-I and ObsID:17857 for ACIS-S). The simulation was done for 50 iterations in order to obtain enough counts to extract a high S/N PSF\footnote{http://cxc.harvard.edu/ciao/PSFs/chart2/runchart.html}. Then the {\sc MARX} (v5.3) software and {\sc dmmerge} script were used to create a combined pseudo event file from all the 50 iterations. Finally, a high S/N PSF radial profile was extracted from the combined event file. This procedure was repeated for ACIS-I and ACIS-S at 3.3, 5.0 and 7.0 keV separately, because \axj\ was at slightly different off-axis angles during these two observation modes. Note that in the simulations for ACIS-S, there was no typical `X' pattern from the HETG arms, thus the simulated PSF profile is not affected by the gratings.

Meanwhile, Gaetz (2010) performed a detailed analysis of the PSF wing of HRMA using a deep calibration observation of the low mass X-ray binary Her X-1 (ObsID: 3662), which is bright and has low $N_{H,sca}$ in the LOS. This observation was also used in some other works to measure the PSF profile (e.g. Smith, Edgar \& Shafer 2002; Xiang, Lee \& Nowak 2007; Xiang et al. 2011). Because Her X-1 was heavily piled up in the PSF core, we only used the best-fit King profile in Gaetz (2010) to produce the PSF from 10 arcsec to 500 arcsec. Her X-1 was at a 0.75 arcmin off-axis angle during this observation, while \axj\ was on-axis in the ACIS-S mode and at 1.45 arcmin off-axis in the ACIS-I mode, but the small difference of the off-axis angle should not affect the PSF wing significantly (see the \cxo\ website\footnote{http://cxc.harvard.edu/ciao/PSFs/chart2/caveats.html} and Gaetz 2010).

Since the PSF from the {\sc ChaRT} simulation for $E\ge$ 2 keV may under-predict the flux in the PSF wing\footnotemark[2], while the PSF from Gaetz (2010) was affected by severe pile-up within 10 arcsec, we decided to use the core of the {\sc ChaRT} PSF and the wing of the Gaetz (2010) PSF by matching them within 10-15 arcsec, so that an accurate PSF profile was derived. This final PSF profile was renormalized and adopted in our halo profile modelling.

\begin{figure}
\centering
\includegraphics[bb=200 0 650 610, scale=0.275]{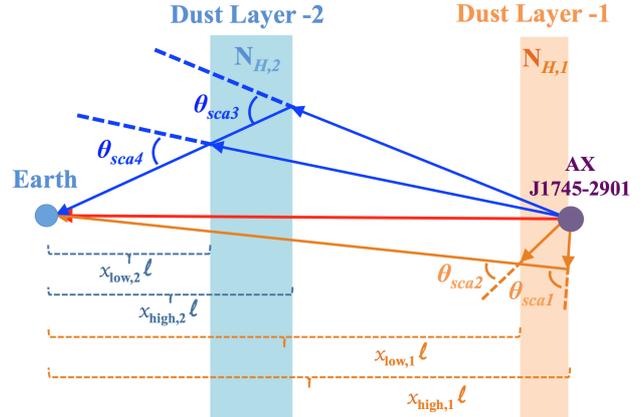} 
\caption{The two dust-layer scenario for the radial profile fitting in Fig.\ref{fig-radfit}. $x_{low,1,2}$ and $x_{high,1,2}$ are the fractional distances of the two dust layers from Earth. $N_{H,1}$ and $N_{H,2}$ are the $N_{H,sca}$ in Layer-1 and Layer-2, separately. $\ell$ is the absolute distance of \axj from Earth. $\theta_{sca1,2}$ and $\theta_{sca3,4}$ show the change of dust scattering angle at a specific viewing angle within the two dust layers.}
\label{fig-cartoon3}
\end{figure}

\subsection{PSF of \xmm\ EPIC-pn}
The on-axis half energy width of EPIC-pn is 16.6 arcsec (\xmm\ Users Handbook\footnote{http://www.cosmos.esa.int/web/xmm-newton/documentation}), implying a much bigger impact on the halo profile than in \cxo. Ghizzardi (2002) studied 110 \xmm\ observations to calibrate the EPIC-pn PSF and provided equations (based on the best-fit King profile) to calculate the 1-dimensional (1D) PSF profile (also see Costantini, Freyberg \& Predehl 2005). Meanwhile, the 1D and 2D PSF profiles were modelled separately with a 1D King profile and a 2D ELLBETA model by the \xmm\ calibration team (Read et al. 2011), with the best-fit parameters given in the latest Current Calibration Files (CCF). However, it was noticed that the 2D ELLBETA model requires further calibration in the PSF wing (Schartel, private communication), while the CCF files only provide 1D PSF parameters for a discrete set of off-axis angles and energies, so we decided to adopt the Ghizzardi (2002) PSF which should be accurate enough for our study. Note that the off-axis angle of \axj\ is 1.454-1.678 arcmin in all the EPIC-pn observations, and the PSF changes very little within these small off-axis angles.

\section{Modelling the Dust Scattering Halo of \axj}
\label{rp-model}

\subsection{Radial Profile Fitting of the Dust Scattering Halo}
\label{sec-radfit}
The shape of the dust scattering halo depends on the dust grain model (see Section~\ref{sec-grainmodel}). In Fig.\ref{fig-haloshape}, we compare the halo profiles using some representative dust grain models, including the classic MRN77 dust grain model, two dust grain models from WD01 with [$R_{V}=3.1,~Bc=6.0,~Case-A$] (adopted by Valencic \& Smith 2015 as the standard WD01 dust grain model) and [$R_{V}=5.5,~Bc=3.0,~Case-B$] (highest intensity in the halo core) where $Bc$ ($\times10^{5}$) is the carbon abundance per hydrogen nucleus, and three ZDA04 dust grain models. In this paper, we tried all dust grain models in MRN77, WD01, ZDA04 and XLNW to do the halo profile modelling and compared the results.

A reasonable dust scattering model should reproduce not only the halo radial profile at a specific energy, but also the energy dependence of the halo. Besides, the halo profiles from \cxo\ and \xmm\ should contain the same intrinsic halo shape except for the different instrumental PSF. Therefore, we decided to fit the halo profiles from \cxo\ ACIS-I and \xmm\ in the 2-4, 4-6 and 6-10 keV bands, simultaneously, i.e. six radial profiles in total (see Fig.\ref{fig-radfit}). The dust scattering calculation is computationally intensive (Smith, Valencic, Corrales 2016; Corrales \& Paerels 2015). In order to increase the computing efficiency, we calculated the spectral weighted effective energy within the 2-4, 4-6 and 6-10 keV bands as 3.3, 5.0 and 7.0 keV and fixed the photon energy at these values in the model. This should not affect the accuracy because the halo shape is not sensitive to a small change of the effective energy. Assuming a rotational symmetry, we performed 2D convolution for the instrumental PSF in order to fit the halo profile observed by various instruments. All the fittings were performed with the {\sc Sherpa} fitting engine provided in the {\sc CIAO} (v4.8.2) software\footnote{http://cxc.cfa.harvard.edu/ciao}.

\subsubsection{One Dust-Layer Model}
\label{sec-radfit1}
First, we assumed a point source and one foreground dust layer of smooth distribution (as shown in Fig.\ref{fig-cartoon1}). The lower and upper boundary of the dust layer and the $N_{H,sca}$ were all free parameters. The incident flux to the dust layer was linked to the flux of the point source, which was also a free parameter because the fraction of scattering flux in the total observed flux was unknown. Only single scattering was considered (see Section~\ref{sec-multiscat} for the discussion of multiple scattering). The absolute distance of \axj\ does not affect the halo profile fitting because all distances in the model are fractional. Fig.\ref{fig-1dl} shows an example of the one-layer model fit to the 4-6 keV radial profile of \axj\ observed by \cxo\ ACIS-I, using the COMP-AC-S dust grain model. The best-fit $\chi^2=$ 1329 for 178 degrees of freedom (dof) for all six radial profiles. The fitting also required the dust layer to extend from Earth to the source with $N_{H,sca}=1.7\times10^{23}~cm^{-2}$. It is clear that this model is too simple to reproduce the halo profile around \axj. We tried other dust grain models but none could produce a reasonably good fit with reduced $\chi^2_v<2$. The residuals in Fig.\ref{fig-1dl} clearly indicate that the curvature of the radial profile is more complex than the dust scattering halo from one smooth dust layer regardless of the dust grain model used.

\begin{figure}
\centering
\includegraphics[bb=40 120 540 648, scale=0.43]{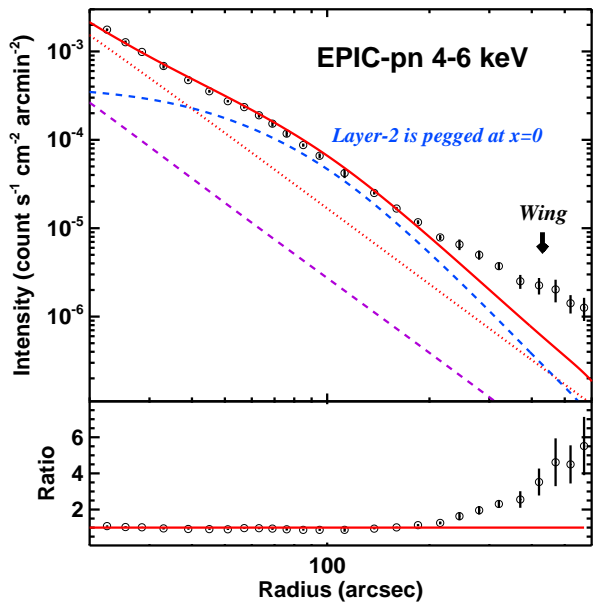} 
\caption{Radial profile observed by \xmm\ EPIC-pn in the 4-6 keV band, overplotted with the EPIC-pn PSF (red dotted curve), dust Layer-1 (magenta dashed curve), Layer-2 whose fractional distance is pegged at 0 and fractional width pegged at the lower limit of 0.01 (blue dashed curve), and the total model (red solid curve). A significant excess flux exists outside 200 arcsec which cannot be fitted by any dust model (black arrow).}
\label{fig-wing}
\end{figure}

\begin{figure*}
\centering
\begin{tabular}{ccc}
\includegraphics[bb=50 120 540 648, scale=0.32]{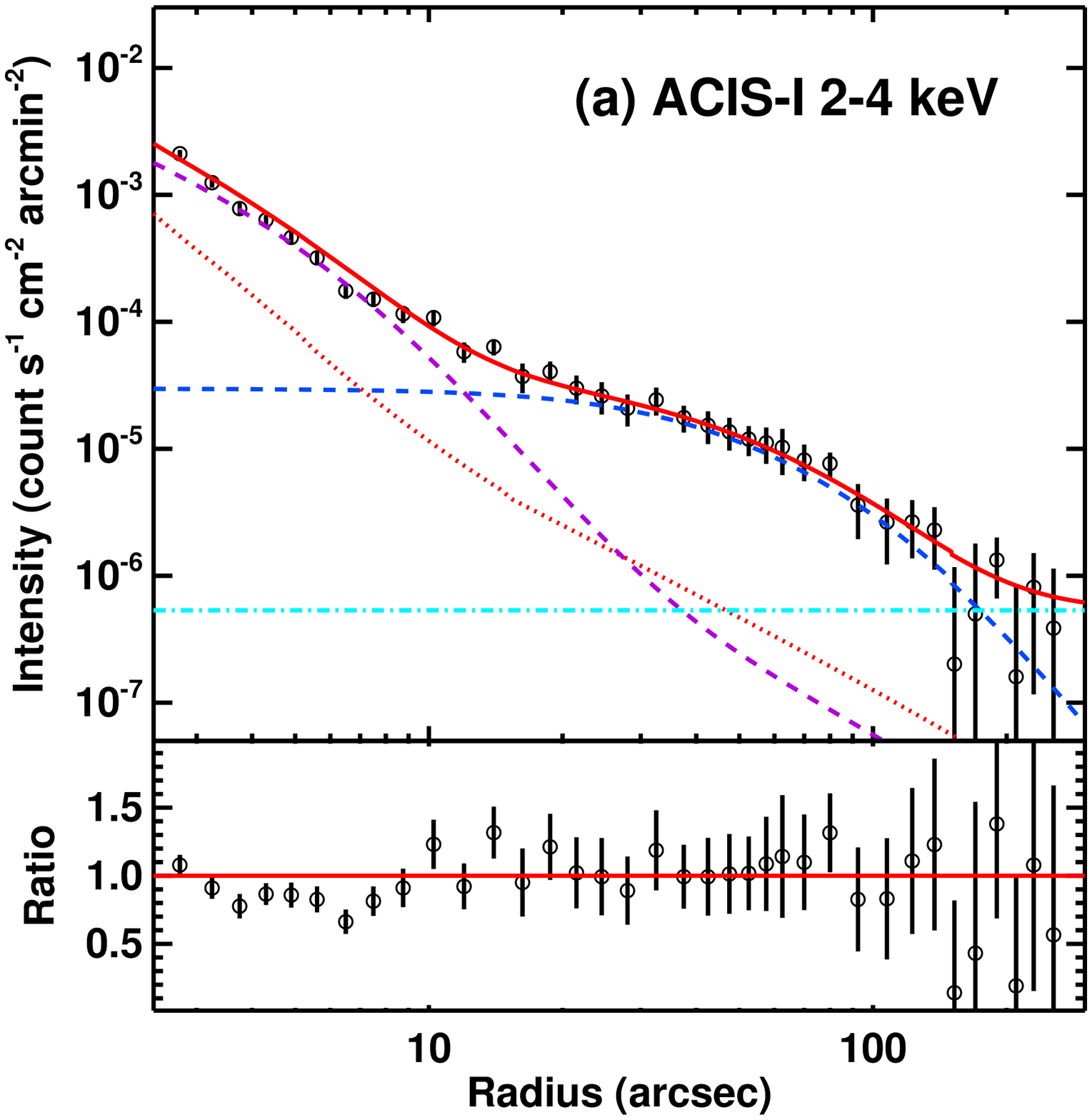} &
\includegraphics[bb=50 120 540 648, scale=0.32]{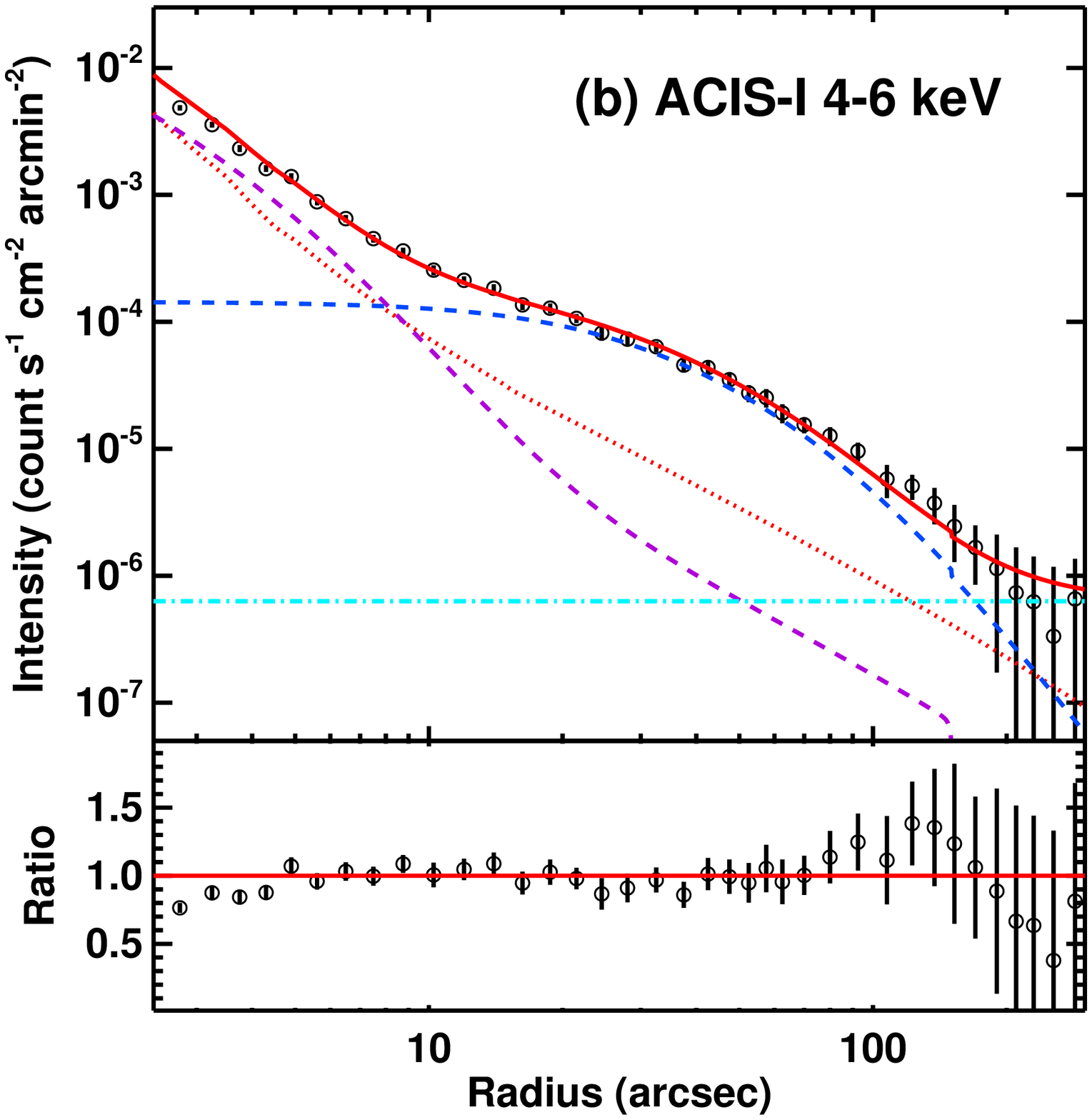} &
\includegraphics[bb=50 120 540 648, scale=0.32]{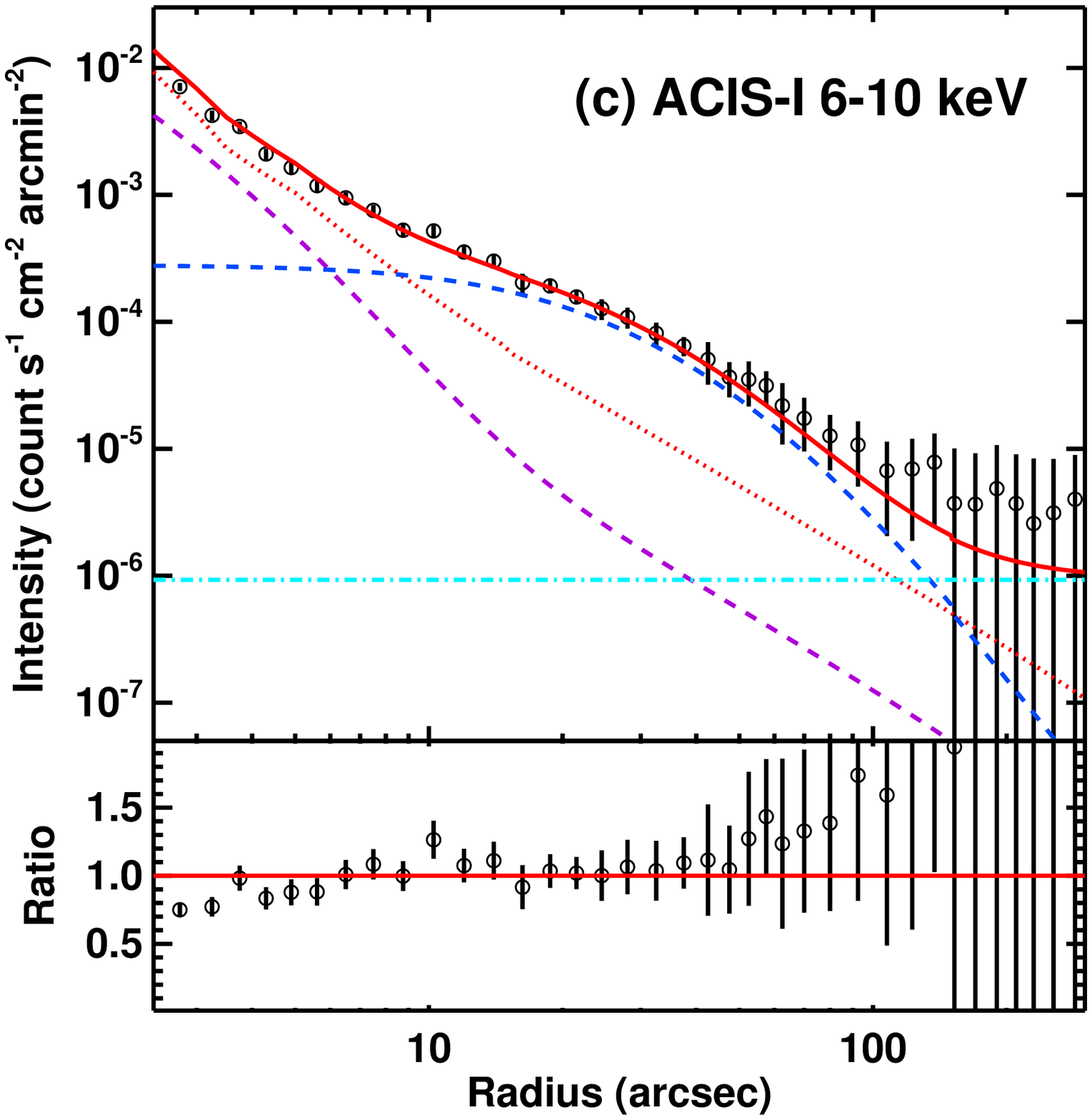} \\
\includegraphics[bb=50 120 540 648, scale=0.32]{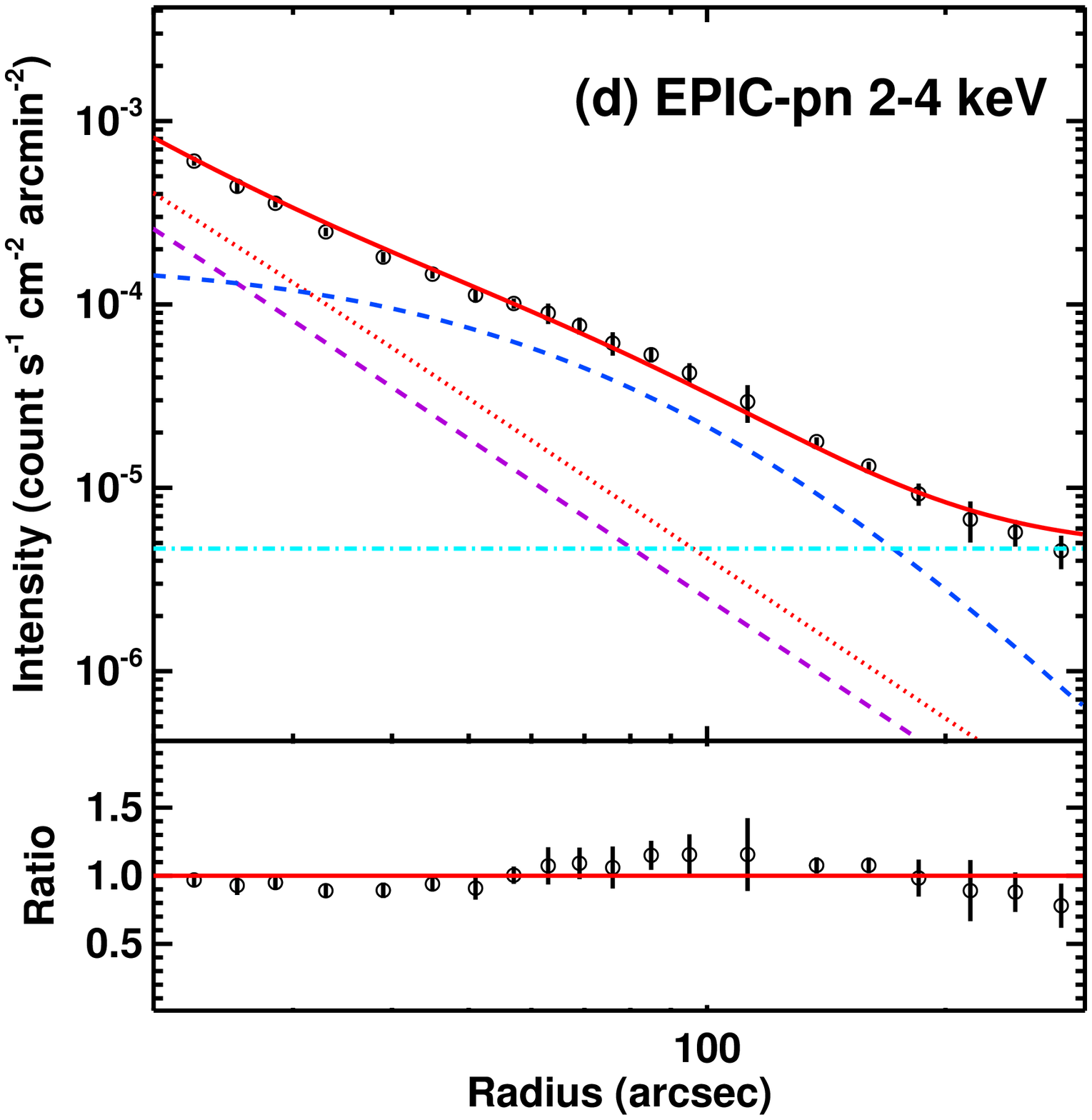} &
\includegraphics[bb=50 120 540 648, scale=0.32]{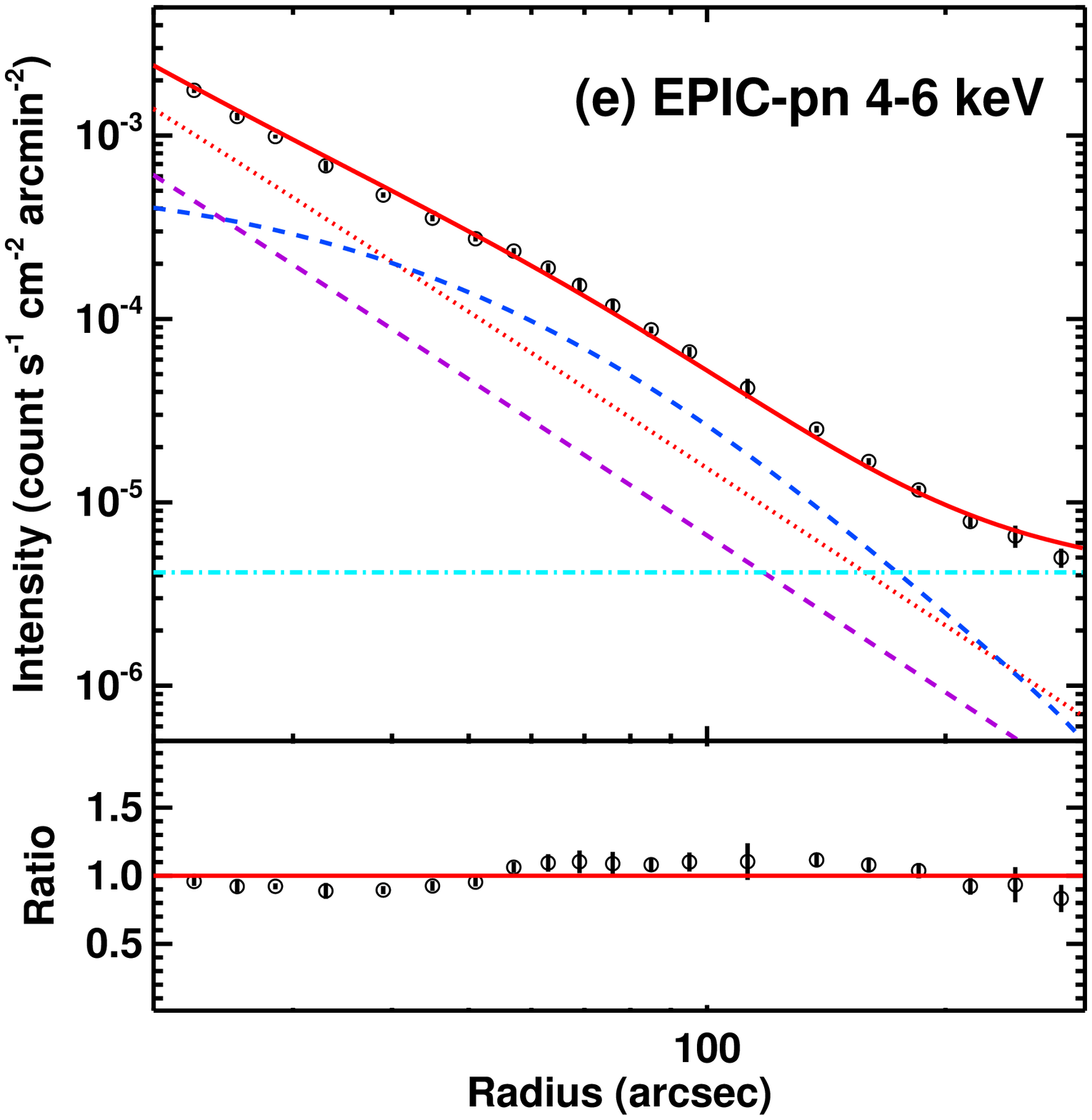} &
\includegraphics[bb=50 120 540 648, scale=0.32]{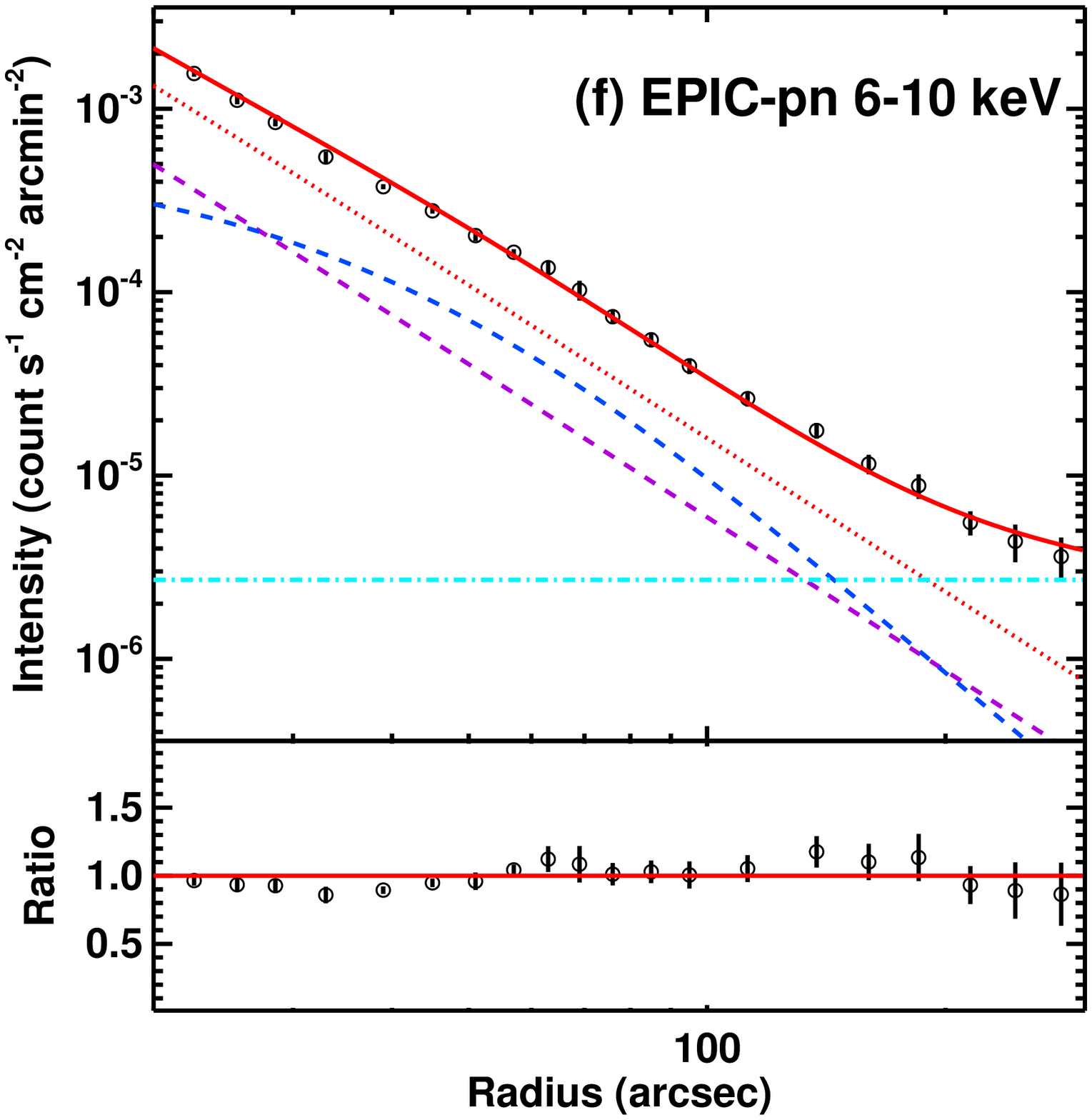} \\
\end{tabular}
\caption{The radial profile fitting for \cxo\ ACIS and \xmm\ EPIC pn in the 2-4, 4-6 and 6-10 keV bands using two dust layers with the COMP-AC-S dust grain model (ZDA04). The background is already subtracted (see Section~\ref{sec-obs-cxo}). The red solid curve shows the total model. The red dotted curve shows the instrumental PSF. The two dashed curves show the dust scattering from the two dust layers (magenta: Layer-1; blue: Layer-2). The cyan line is a flat component to account for the halo wing dominating outside $\sim$200 arcsec, which cannot be reproduced directly by any dust grain model (see Section~\ref{sec-wing}). The best-fit parameters can be found in Table~\ref{tab-radfit}. The radial profiles before background subtraction can be found in Fig.\ref{app-fig-radprof}.}
\label{fig-radfit}
\end{figure*}

\subsubsection{Two Dust-Layers without Wing}
\label{sec-radfit2-nowing}
Then we adopted a two-layer model, i.e. Layer-1 + Layer-2. Fig.\ref{fig-cartoon3} shows the setup and free parameters of the two dust layers, including the $N_{H,sca}$, lower and upper boundaries of each layer. Layer-1 was closer to \axj\ and layer-2 was closer to Earth, and so the halo from Layer-2 was more extended than that from Layer-1. The parameters of these two layers were independent from each other. As a first order consideration, we assumed the same dust grain model for the two layers. This two-layer model improved the halo fitting significantly. For example, we found $\chi^2=$ 664 for 175 dof in the case of COMP-AC-S dust grain model. Fig.\ref{fig-wing} shows that this two-layer model cannot fit the halo wing $\ge$ 200 arcsec observed by \xmm. The discrepancy between the the wing and the best-fit model increases towards large radii up to a factor of 6 at 600 arcsec. The total flux in the detected halo wing (200-600 arcsec) is a factor of $6.0\pm1.0$ higher than in the model. We emphasise that the fractional distance of layer-2 in Fig.\ref{fig-wing} (blue dashed curve) has already pegged at 0 with the fractional width reaching the lower limit of 0.01, indicating that this wing component cannot be solely explained by the distance of the layer\footnote{we chose 0.01 to avoid numerical problem in the halo calculation. Changing it to an even smaller value causes negligible difference to the halo profile.}. We find that this wing component is related to the X-ray emission from \axj\ and rises from some dust grains with relatively small sizes (see Section~\ref{sec-wing}), while scattering from large dust grains still dominates at small radii.

\subsubsection{Two Dust-Layers with Wing}
\label{sec-radfit2}
Because of the halo wing component detected, we restricted our fitting to the radial profile less than 300 arcsec and added a free constant parameter to the model to account for the wing above 200 arcsec. Since this wing component should exist in both \cxo\ and \xmm\ observations, we assumed that it had the same scaling factor to the source flux in both instruments. Therefore, the final two-layer model was Layer-1 + Layer-2 + Wing.

Based on this new two-layer model, we fitted the halo profile with various dust grain models. Fig.\ref{fig-radfit} shows the halo fitting for the COMP-AC-S dust grain model. 18 out of 19 dust grain models produce $\chi^2_{v}<2$, indicating a reasonably good fit. The best $\chi^2_{v}=1.29$ was found by the BARE-GR-B dust grain model (Table~\ref{tab-radfit}). The best-fit parameters and their statistical error are listed in Table~\ref{tab-radfit} and \ref{tab-tau}. It is clear that the statistical errors are much smaller than the dispersion of best-fit values among all the dust grain models, indicating significant systematic uncertainty due to the assumptions made in the dust grain model. Other systematics include the assumptions about the uniform dust distribution inside every layer, the same dust grain model for all the foreground dust, Gaussian approximation for the form factor, calibration systematics especially in the PSF wing.

Among all the 19 dust grain models, the highest $N_{H,sca}=(29.7\pm1.3)\times10^{22}~cm^{-2}$ was found for the COMP-AC-B dust grain model, while the lowest $N_{H,sca}=(5.0\pm0.2)\times10^{22}~cm^{-2}$ was found for the WD01-B dust grain model. This large $N_{H,sca}$ dispersion was also found in previous works (e.g. Smith, Edgar \& Shafer 2002; Valencic \& Smith 2008; Valencic et al. 2009;  Xiang et al. 2011), which is mainly due to different assumptions about the dust-to-gas ratio, abundances and size distribution made in different dust grain models. However, the mean best-fit $N_{H,sca}~=~16.9\times10^{22}~cm^{-2}$ and the COMP-AC-S best-fit $N_{H,sca}~=~19.2\times10^{22}~cm^{-2}$ are still roughly consistent with $N_{H,abs}\sim2.0\times10^{23}~cm^{-2}$ using the AG89 solar abundances (Ponti et al. 2015b).

For all dust grain models, the fitting requires dust Layer-1 to be close to \axj. Depending on the dust grain model used, the fractional distance of Layer-1 is $\lesssim$ 0.1 from the source and containing (19-34)\%\ of the total dust along the LOS. Layer-2 is more extended and contains (66-81)\%\ of the intervening dust, with its lower boundary always pegged at 0, indicating that part of this layer is very close to Earth. From Table~\ref{tab-radfit}, we see that the upper boundary of layer-2 ($x_{high,2}$) also depends on the dust grain model, ranging from 0.01 to 0.90. The lowest values are found in the COMP-NC-B ($x_{high,2}=0.01$) and COMP-NC-GF ($x_{high,2}=0.11$) dust grain models, but their $\chi^2$ are also the largest. The highest value is found in the BARE-GR-B model ($x_{high,2}=0.90$). We notice that the boundary of the two layers mainly depend on the size distribution of dust grains in each model. A higher fraction of small dust grains will produce a more extended halo profile, and so $x_{high,2}$ ($x_{low,1}$) will increase (decrease) to enhance the halo intensity at small radii in order to fit the observed halo profile. The mean $x_{high,2}$ is 0.64, which is about half way towards \axj. The separation between these two layers is quite significant compared to the statistical error, although the value ranges from 0.09 to 0.88 among all the dust grain models (Table~\ref{tab-tau}). Adding more dust layers can improve the fitting statistics but also introduce more parameter degeneracies. To better resolve these two major dust layers, it is necessary to consider the timing properties of the halo due to the variability of \axj\ (e.g. the eclipsing signal), which will be presented in a separate work (Jin et al. in prep.).

\begin{figure}
\centering
\includegraphics[bb=50 216 558 669,scale=0.46]{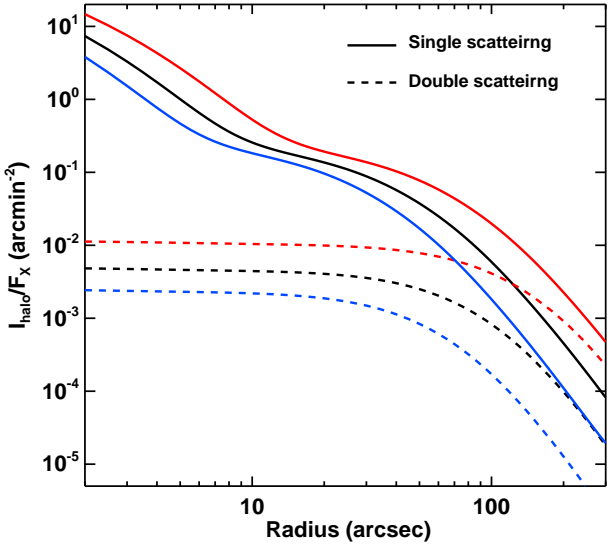}
\caption{Comparison of the radial profiles from the single scattering (solid lines) and double scattering (dash lines) for 3.3 keV (red, $\tau=0.82$), 5.0 keV (black, $\tau=0.55$) and 7.0 keV (blue, $\tau=0.42$), based on the two-layer model with COMP-AC-S dust grains (Table~\ref{tab-tau}).}
\label{fig-2scat}
\end{figure}

\subsection{Optical Depth and Multiple Scattering}
\label{sec-multiscat}
Although our two-layer single scattering model can provide a reasonably good fit to the halo profile of \axj\ in all three energy bands, it is still necessary to check if multiple scattering is important. The inclusion of higher order scattering will produce a brighter and more extended halo than single scattering (Mathis \& Lee 1991). Multiple scattering starts to dominate when the scattering optical depth ($\tau_{sca}$) is significantly bigger than unity (e.g. Mathis \& Lee 1991; Xiang, Lee \& Nowak 2007). Since \axj\ is an eclipsing source, the total flux and halo flux can be measured directly from inside and outside the eclipsing phase, then $\tau_{sca}$ can be calculated using Eq.\ref{equ-tau}.

In order to measure the flux accurately, we chose a different \xmm\ observation (ObsID: 0723410301) where \axj\ was not piled up. The same data reduction procedures were applied to this observation. Source light curves were extracted from a circular region of 200 arcsec to include the dust scattering halo in the 2-4, 4-6 and 6-10 keV bands. The remaining halo flux outside 200 arcsec should be negligible (see Fig.\ref{fig-radfit}). From background subtracted light curves, $F_{sca}/F_{obs}$ was found to be 0.57, 0.32, 0.16 for the three energy bands, implying $\tau_{sca}=$ 0.84, 0.39 and 0.17. We also calculated $\tau_{sca}$ from every best-fit two-layer model, which are consistently less than 1 (Table~\ref{tab-tau}). Similar results can be found using common relations between $\tau_{sca}$, $N_{H,abs}$ and the V band extinction $A_{V}$ (see Section~\ref{sec-dgratio}). These scattering optical depths indicate that single scattering should dominate in the case of \axj. As a further check, we followed the equations in Mathis \& Lee (1991) and Xiang, Lee \& Nowak (2007) to calculate the halo profile from double scattering using the best-fit COMP-AC-S two-layer model. Fig.\ref{fig-2scat} shows that the intensity of the double scattering component is indeed dominated by the single scattering component. Therefore, we conclude that it is sufficient to consider only single scattering in this work.

\begin{table*}
 \centering
  \begin{minipage}{178mm}
  \centering
   \caption{Best-fit parameters for all the 19 dust grain models using the {\it Two Dust-Layers with Wing} model. $x_{high/low,1/2}$ and $N_{H,1/2}$ are the fractional distances and $N_{H,sca}$ of the two dust layers, as shown in Fig.\ref{fig-cartoon3}. The COMP-AC-S dust grain model (indicated by ${\dagger}$) was recommended for the GC direction by Fritz et al. (2011). $l$: the parameter pegs at the lower limit. $\chi^2/dof$ (ALL) is for the simultaneous fitting to all the six radial profiles in Fig.\ref{fig-radfit}, while $\chi^2/dof$ (CXO) is the fitting statistics of the three \cxo\ radial profiles. Error bars are calculated for the 90\% confidence range. We emphasise that systematic uncertainties are more important, such as the scatter of best-fit parameters between different dust grain models.}
     \begin{tabular}{llccccccccc}
     \hline
     No. & Dust Model& $x_{low,2}$ & $x_{high,2}$  & $x_{low,1}$ & $x_{high,1}$ & $N_{H,2}$ & $N_{H,1}$ & $\chi^2/dof$ & $\chi^2/dof$\\
        & &  &  & & & ($10^{22}~cm^{-2}$) & ($10^{22}~cm^{-2}$) & (ALL) & (CXO)\\
\hline
     0 & MRN77                          &0$^{l}$ &0.8949 $^{+0.0138}_{-0.0716}$ & 0.9783 $^{+0.0010}_{-0.0021}$&0.9976 $^{+0.0008}_{-0.0004}$ & 13.2 $^{+1.1}_{-1.1}$ &6.8 $^{+0.9}_{-0.9}$ & 199/131& 97/83\\
     1 &BARE-GR-S		         & 0$^{l}$ &0.8092 $^{+0.0082}_{-0.0643}$ &0.9553 $^{+0.0022}_{-0.0062}$&0.9975 $^{+0.0017}_{-0.0030}$ & 9.5 $^{+0.5}_{-0.5}$ &2.3 $^{+0.3}_{-0.3}$& 204/131& 126/83\\
     2 &BARE-GR-FG		 & 0$^{l}$ &0.8314 $^{+0.0131}_{-0.0472}$ &0.9611 $^{+0.0034}_{-0.0084}$&0.9989 $^{+0.0011}_{-0.0006}$& 10.5 $^{+0.8}_{-0.8}$ &3.0 $^{+0.5}_{-0.5}$ & 198/131& 107/83\\
     3 &BARE-GR-B                  & 0$^{l}$ &0.9010 $^{+0.0046}_{-0.0421}$ &0.9760 $^{+0.0024}_{-0.0087}$&0.9984 $^{+0.0011}_{-0.0003}$&15.1 $^{+0.9}_{-0.9}$ &4.8 $^{+0.5}_{-0.5}$  & 169/131& 94/83\\
     4 &COMP-GR-S                 & 0$^{l}$ &0.6242 $^{+0.0035}_{-0.0423}$ &0.9403 $^{+0.0102}_{-0.0026}$ &0.9985 $^{+0.0014}_{-0.0001}$&12.0 $^{+0.8}_{-0.8}$ &5.2 $^{+0.5}_{-0.5}$  & 230/131& 100/83\\
     5 &COMP-GR-FG              & 0$^{l}$ &0.7465 $^{+0.0117}_{-0.0209}$ &0.9523 $^{+0.0076}_{-0.0036}$&0.9976 $^{+0.0011}_{-0.0001}$&11.6 $^{+0.7}_{-0.7}$ &4.9 $^{+0.5}_{-0.5}$  & 229/131& 97/83\\
     6 &COMP-GR-B                 & 0$^{l}$ &0.7913 $^{+0.0116}_{-0.0152}$ &0.9557 $^{+0.0040}_{-0.0070}$&0.9986 $^{+0.0008}_{-0.0006}$&11.3 $^{+0.8}_{-0.8}$ &3.2 $^{+0.5}_{-0.5}$  & 207/131& 113/83\\
     7 &BARE-AC-S		         & 0$^{l}$ &0.8491 $^{+0.0143}_{-0.0193}$&0.9637 $^{+0.0039}_{-0.0088}$&0.9976 $^{+0.0023}_{-0.0005}$&9.6 $^{+0.6}_{-0.6}$ & 3.2 $^{+0.3}_{-0.3}$  & 196/131& 101/83\\
     8 &BARE-AC-FG                & 0$^{l}$ &0.8394 $^{+0.0218}_{-0.0172}$ &0.9625 $^{+0.0040}_{-0.0089}$ &0.9977 $^{+0.0022}_{-0.0004}$&9.3 $^{+0.5}_{-0.5}$ &3.1 $^{+0.3}_{-0.3}$  & 198/131& 104/83\\
     9 &BARE-AC-B		         & 0$^{l}$ &0.8783 $^{+0.0283}_{-0.0561}$  &0.9665 $^{+0.0017}_{-0.0088}$ &0.9982 $^{+0.0009}_{-0.0010}$ &11.3 $^{+0.6}_{-0.6}$ &3.0 $^{+0.3}_{-0.3}$  & 184/131& 109/83\\
   10 &COMP-AC-S$^{\dagger}$  & 0$^{l}$ &0.5143 $^{+0.0092}_{-0.5143}$ &0.9389 $^{+0.0060}_{-0.0208}$ &0.9978 $^{+0.0009}_{-0.0001}$ &13.4 $^{+1.3}_{-1.3}$ &5.8 $^{+0.8}_{-0.8}$  & 224/131& 103/83\\
   11 &COMP-AC-FG                  &0$^{l}$&0.6737 $^{+0.0067}_{-0.2864}$ &0.9474 $^{+0.0040}_{-0.0180}$&0.9977 $^{+0.0009}_{-0.0001}$&12.6 $^{+0.8}_{-0.8}$ &5.4 $^{+0.5}_{-0.5}$  &239/131& 105/83\\
   12 &COMP-AC-B                  &0$^{l}$&0.2954 $^{+0.0025}_{-0.2954}$ &0.9225 $^{+0.0127}_{-0.0175}$&0.9976 $^{+0.0018}_{-0.0009}$&23.3 $^{+1.1}_{-1.1}$ &6.4 $^{+0.7}_{-0.7}$  &213/131& 137/83\\
   13 &COMP-NC-S                  &0$^{l}$ &0.3356 $^{+0.0059}_{-0.3356}$ &0.9306 $^{+0.0025}_{-0.0358}$&0.9979 $^{+0.0012}_{-0.0002}$&15.5 $^{+1.0}_{-1.0}$ &6.5 $^{+0.7}_{-0.7}$  & 234/131& 121/83\\
   14 &COMP-NC-FG                  &0$^{l}$&0.1093 $^{+0.0125}_{-0.1093}$ &0.9071 $^{+0.0072}_{-0.0234}$&0.9991 $^{+0.0004}_{-0.0002}$ &16.2 $^{+1.3}_{-1.3}$ &5.4 $^{+0.7}_{-0.7}$  &211/131& 136/83\\
   15 &COMP-NC-B                  &0$^{l}$&0.0100 $^{+0.0273}_{l}$ &0.8873  $^{+0.0161}_{-0.0194}$ &0.9990 $^{+0.0005}_{-0.0001}$ &17.9 $^{+1.2}_{-1.2}$ &5.9 $^{+0.7}_{-0.7}$ &271/131& 169/83\\
   16 &WD01-A                                 &0$^{l}$ &0.8302  $^{+0.0086}_{-0.1314}$&0.9628 $^{+0.0039}_{-0.0239}$&0.9975 $^{+0.0017}_{-0.0001}$&8.4 $^{+0.5}_{-0.5}$ &4.5 $^{+0.3}_{-0.3}$ & 246/131& 97/83\\
   17 &WD01-B                                 &0$^{l}$ &0.4035 $^{+0.0122}_{-0.4035}$&0.9326  $^{+0.0027}_{-0.0243}$&0.9977 $^{+0.0011}_{-0.0001}$&3.6 $^{+0.2}_{-0.2}$ &1.4 $^{+0.1}_{-0.1}$ & 235/131& 116/83\\
   18 &XLNW 					&0$^{l}$ &0.8772 $^{+0.0034}_{-0.0517}$&0.9701 $^{+0.0088}_{-0.0065}$&0.9974 $^{+0.0019}_{-0.0004}$&11.6 $^{+0.6}_{-0.6}$ &4.3 $^{+0.4}_{-0.4}$ & 175/131& 88/83\\
\hline
    \multicolumn{2}{c}{{\it min. value}} & 0$^{l}$ & 0.0100$^{l}$ & 0.8873 & 0.9974 & 3.6 & 1.4 & -- & --\\
    \multicolumn{2}{c}{{\it max. value}} & 0$^{l}$ & 0.9010 & 0.9783 & 0.9991 & 23.3 & 6.8 & -- & --\\
    \multicolumn{2}{c}{{\it mean value}}  & 0$^{l}$ & 0.6429 & 0.9479 & 0.9980 & 12.4  & 4.5 & -- & --\\
    \multicolumn{2}{c}{{\it median value}} & 0$^{l}$ & 0.7913 & 0.9523 & 0.9977 & 11.6 & 4.5 & -- & --\\
     \hline
        \end{tabular}
     \label{tab-radfit}
 \end{minipage}
\end{table*}

\begin{table*}
\centering
\begin{minipage}{178mm}
\centering
   \caption{Parameters calculated from the best-fit models in Table~\ref{tab-radfit} with propagated statistical errors. $\tau$ is the dust scattering optical depth calculated from each of the best-fit models. $f_{N_{H,2}}$ $= N_{H,2}/(N_{H,1}+N_{H,2})\times100\%$, where $N_{H,1}$ and $N_{H,2}$ are the $N_{H,sca}$ in Layer-1 and Layer-2, separately. $f_{N_{H,2}}$ indicates the percentage of dust contained in Layer-2 along the LOS.}
     \begin{tabular}{llcccccc}
     \hline
     No. & Dust Model & $x_{low,1}-x_{high,2}$ & $N_{H,2}+N_{H,1}$ & $f_{N_{H,2}}$ & $\tau (2-4 keV)$ & $\tau (4-6 keV)$ & $\tau (6-10 keV)$\\
      & & & ($10^{22}~cm^{-2}$)& ($\%$) & & & \\
     \hline
    0 & MRN77 & 0.0834 $^{+0.0716}_{-0.0140}$ & 20.0 $^{+1.4}_{-1.4}$ & 66.0 $^{+3.5}_{-3.5}$ & 0.80 & 0.53 & 0.41 \\
    1 & BARE-GR-S & 0.1461 $^{+0.0643}_{-0.0103}$ &11.8 $^{+0.6}_{-0.6}$ & 80.5 $^{+2.2}_{-2.2}$ & 0.51 & 0.28 & 0.18\\
    2 & BARE-GR-FG & 0.1297 $^{+0.0473}_{-0.0156}$ & 13.5 $^{+0.9}_{-0.9}$ & 77.8 $^{+3.2}_{-3.2}$& 0.65 & 0.42 & 0.32\\
    3 & BARE-GR-B & 0.0750 $^{+0.0421}_{-0.0098}$ & 19.9 $^{+1.0}_{-1.0}$ & 75.9 $^{+2.2}_{-2.2}$& 0.66 & 0.40 & 0.28\\
    4 & COMP-GR-S & 0.3161 $^{+0.0960}_{-0.0044}$ &17.2 $^{+0.9}_{-0.9}$ & 69.8 $^{+2.5}_{-2.5}$& 0.93 & 0.65 & 0.53\\
    5 & COMP-GR-FG & 0.2058 $^{+0.0222}_{-0.0122}$ & 16.5 $^{+0.9}_{-0.9}$ & 70.3 $^{+2.5}_{-2.5}$& 0.79 & 0.51 & 0.37\\
    6 & COMP-GR-B & 0.1644 $^{+0.0157}_{-0.0135}$ & 14.5 $^{+0.9}_{-0.9}$ & 77.9 $^{+3.0}_{-3.0}$& 0.61 & 0.38 & 0.28\\
    7 & BARE-AC-S & 0.1146 $^{+0.0197}_{-0.0168}$ &12.8 $^{+0.7}_{-0.7}$ & 75.0 $^{+2.1}_{-2.1}$& 0.63 & 0.37 & 0.26\\
    8 & BARE-AC-FG & 0.1231 $^{+0.0177}_{-0.0235}$ & 12.4 $^{+0.6}_{-0.6}$ & 75.0 $^{+2.1}_{-2.1}$& 0.62 & 0.37 & 0.25\\
    9 & BARE-AC-B & 0.0882 $^{+0.0561}_{-0.0296}$ &14.3 $^{+0.7}_{-0.7}$ & 79.0 $^{+1.9}_{-1.9}$& 0.57 & 0.33 & 0.23\\
    10 & COMP-AC-S$^{\dagger}$ & 0.4246 $^{+0.5143}_{-0.0227}$ & 19.2 $^{+1.5}_{-1.5}$ & 69.8 $^{+3.6}_{-3.6}$& 0.82 & 0.55 & 0.43\\
    11 & COMP-AC-FG & 0.2737 $^{+0.2864}_{-0.0192}$ & 18.0 $^{+0.9}_{-0.9}$ & 70.0 $^{+2.4}_{-2.4}$ & 0.77 & 0.50 & 0.37\\
    12 & COMP-AC-B & 0.6271 $^{+0.2957}_{-0.0177}$ & 29.7 $^{+1.3}_{-1.3}$ & 78.5 $^{+2.0}_{-2.0}$& 0.55 & 0.33 & 0.22\\
    13 & COMP-NC-S & 0.5950 $^{+0.3356}_{-0.0363}$ & 22.0 $^{+1.2}_{-1.2}$ & 70.5 $^{+2.6}_{-2.6}$& 0.78 & 0.53 & 0.41\\
    14 & COMP-NC-FG & 0.7978 $^{+0.1095}_{-0.0265}$ & 21.6 $^{+1.5}_{-1.5}$ & 75.0 $^{+2.9}_{-2.9}$& 0.85 & 0.63 & 0.52\\
    15 & COMP-NC-B & 0.8773 $^{+0.0189}_{-0.0335}$ & 23.8 $^{+1.4}_{-1.4}$ & 75.2 $^{+2.5}_{-2.5}$ & 0.71 & 0.48 & 0.38\\
    16 & WD01-A & 0.1326 $^{+0.1315}_{-0.0254}$ &12.9 $^{+0.6}_{-0.6}$ & 65.1 $^{+2.0}_{-2.0}$ & 1.07 & 0.76 & 0.61\\
    17 & WD01-B & 0.5291 $^{+0.4035}_{-0.0272}$ & 5.0 $^{+0.2}_{-0.2}$ & 72.0 $^{+1.8}_{-1.8}$ & 0.75 & 0.51 & 0.40\\
    18 & XLNW &  0.0929 $^{+0.0524}_{-0.0073}$ & 15.9 $^{+0.7}_{-0.7}$ & 73.0 $^{+2.1}_{-2.1}$ & 0.71 & 0.43 & 0.30\\
     \hline
      \multicolumn{2}{c}{{\it min. value}} & 0.0882 & 5.0 & 66.0 & 0.51 & 0.28 & 0.18\\
      \multicolumn{2}{c}{{\it max. value}} & 0.8773 & 29.7 & 80.5 & 1.07 & 0.76 & 0.61\\
      \multicolumn{2}{c}{{\it mean value}} & 0.3051 & 16.9 & 74.4 & 0.73 & 0.47 & 0.36\\
      \multicolumn{2}{c}{{\it median value}} & 0.1644 & 16.5 & 75.0 & 0.71 & 0.48 & 0.37\\
    \hline
     \end{tabular}
  \label{tab-tau}
\end{minipage}
\end{table*}

\begin{figure}
\centering
\includegraphics[bb=40 190 504 648,scale=0.49]{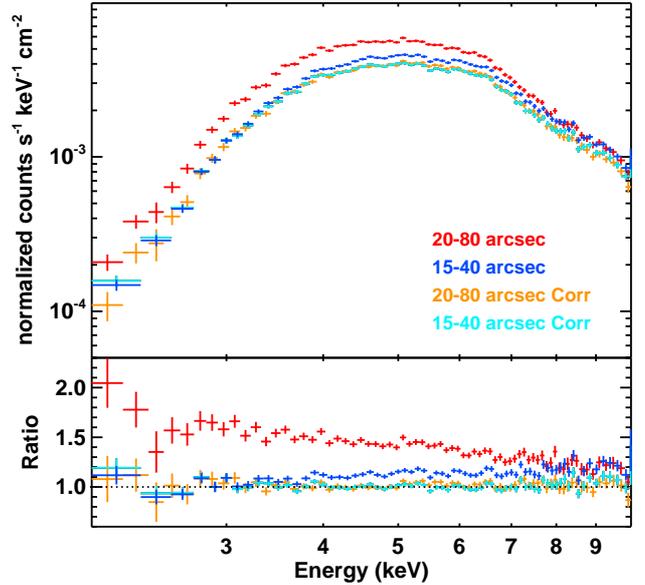}
\caption{\xmm\ EPIC-pn spectra of \axj\ from different extraction regions (ObsID: 0724210201). The red spectrum is from a 20-80 arcsec annulus region, which is both significantly brighter and softer than the blue spectrum from a 15-40 arcsec annulus region. Both spectra have been corrected for the same EPIC-pn PSF. The spectral difference is essentially caused by the dust scattering halo, which can be corrected for using our dust scattering model (i.e. the consistent orange and cyan spectra, see Section~\ref{sec-dscor}). Lower panel shows the ratio of the four spectra to the best-fit absorbed power law model to the orange spectrum.}
\label{fig-specdif}
\end{figure}

\section{Impact of the Dust Scattering for Source Spectra}
\label{sec-spec}
\subsection{Extraction Region Dependence of the Observed Spectrum}
\label{sec-spec1}
X-ray photons passing through the ISM are exposed to both photoelectric absorption and scattering by dust grains (Draine 2003; Ueda et al. 2010). Recent studies on simulated \cxo\ spectra show that the $N_{H,abs}$ can be overestimated by 25\% if the dust scattering opacity is not considered in the spectral fitting (Corrales et al. 2016). Assuming a similar dust-to-gas ratio, the spectral effect of the dust scattering opacity becomes more significant for sources with higher $N_{H,abs}$, such as many sources in the GC.

Dust along the LOS can scatter X-ray photons out of the LOS, which are partially compensated by the photons scattered into the LOS and are subsequently included in the spectral extraction region (see Fig.1 in Smith, Valencic \& Corrales 2016). If an infinite spectral extraction region is adopted, the loss of LOS photons will be fully compensated by the extra photons from the entire dust scattering halo, thereby leaving no dust scattering opacity in the observed spectrum. In reality, the spectral extraction region has a limited size. Besides, a small region of the PSF core is sometimes excised in order to avoid the photon pileup. Therefore, the effective dust scattering opacity is often nonzero and so affects the observed spectrum.

Based on our halo profile modelling, we can investigate this spectral effect for \axj. We extracted \xmm\ pn spectra from an observation (ObsID:0724210201) with two annulus regions (i.e. 15-40 arcsec and 20-80 arcsec) and applied the PSF correction using {\sc SAS} task {\sc arfgen}. Background spectra were extracted from an observation (ObsID:0658600201) where \axj\ was in quiescence. Fig.\ref{fig-specdif} shows the two source spectra (background subtracted) differ significantly from each other in terms of both spectral shape and flux. The spectral difference increases towards soft X-rays, which is consistent with the fact that the effect of dust scattering is stronger for lower energy photons. We note that there could also be spectral discrepancies due to an annulus extraction region for point-like sources because of the PSF calibration issue\footnote{http://www.cosmos.esa.int/web/xmm-newton/sas-watchout-epic-spectra-annular-sources}, but this type of spectral discrepancy is mainly above 4 keV, with $\sim$15\% flux uncertainty in the 5-10 keV band, which is much smaller than what we found in \axj\ due to the dust scattering halo.

To quantify the spectral discrepancy, we fit the two spectra within 2-10 keV using a simple absorbed power law model ({\sc Xspec} model: {\sc tbnew*powerlaw}), with cross-sections of Verner et al. (1996) and WAM00 ISM abundances. For the same $N_{H,abs}$, the best-fit photon index differs by 0.3 (or 6 $\sigma$) between the two spectra, while the flux differs by 30\%, 22\% and 9\% in the 2-4, 4-6 and 6-10 keV bands, respectively (see Table~\ref{tab-specfit}), suggesting the spectral dependence on the source extraction region is very significant and so must be addressed properly.

\begin{figure}
\centering
\includegraphics[scale=0.48]{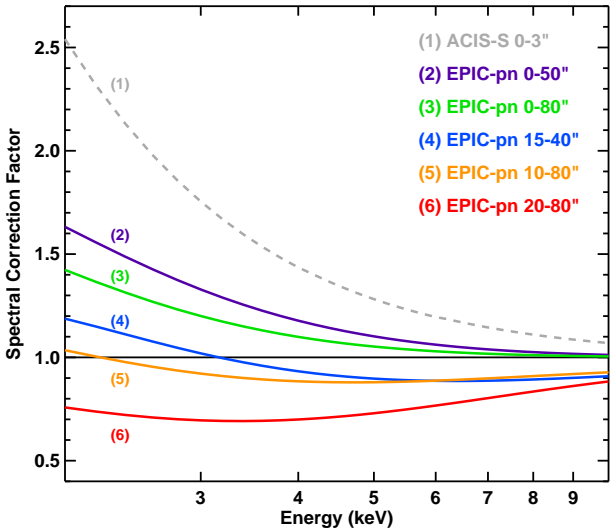}
\caption{Energy dependent spectral correction factor for the dust scattering opacity (Section~\ref{sec-dscor}) for various instruments and spectral extraction regions. Note that these curves are only for the dust on the LOS of \axj.}
\label{fig-specratio1}
\end{figure}

\begin{table*}
\centering
  \begin{minipage}{175mm}
  \centering
 \caption{Comparison of the best-fit parameters for the two spectra in Fig.\ref{fig-specdif} fitted using the {\sc tbnew*powerlaw} model in {\sc Xspec} within 2-10 keV. `Corr' indicates whether the {\sc axjdust} model for the dust scattering opacity correction is incorporated or not. Including the {\sc axjdust} model enables us to obtain consistent spectral parameters from the two very different spectra. $\dagger$ this $N_{H,abs}$ does not necessarily reflect the true $N_{H,abs}$ of \axj, because the absorbed power law model is not a good model for the entire 2-10 keV band (see Section~\ref{sec-dscor}).}
 \begin{tabular}{ccccccccc}
 \hline
 Spectrum & Corr. & $^{\dagger}N_{H,abs}$ & $\Gamma$ & Norm & F$_{2-4 keV}$ & F$_{4-6 keV}$ & F$_{6-10 keV}$ & $\chi^2/dof$ \\
 & & ($10^{22}~cm^{-2}$) & & & \multicolumn{3}{c}{($10^{-12}~erg~cm^{-2}~s^{-1}$)} & \\
 \hline
Spec (15-40\arcsec) & no & 40.9 ${\pm 0.6}$& 3.06 ${\pm 0.04}$ & 1.65 ${\pm 0.1}$ & 15.2 ${\pm 0.1}$ & 68.4 ${\pm 0.4}$ & 97.8 ${\pm 0.5}$ & 1866/1359\\  
Spec (20-80\arcsec) & no & 40.5 ${\pm 0.4}$& 3.38 ${\pm 0.04}$ & 3.46 ${\pm 0.3}$ & 21.8 ${\pm 0.1}$ & 86.3 ${\pm 0.5}$ & 107.3 ${\pm 0.6}$ & 1916/1409\\                     
 \hline
 Spec (15-40\arcsec) & yes & 38.8 ${\pm 0.6}$& 3.01 ${\pm 0.04}$ & 1.30 ${\pm 0.12}$ & 14.1 ${\pm 0.1}$ & 60.7 ${\pm 0.3}$ & 88.6 ${\pm 0.5}$ & 1827/1359\\
 Spec (20-80\arcsec) & yes & 39.1 ${\pm 0.6}$& 3.06 ${\pm 0.04}$ & 1.46 ${\pm 0.13}$ & 14.5 ${\pm 0.1}$ & 62.3 ${\pm 0.3}$ & 89.5 ${\pm 0.5}$ & 1942/1409\\
 \hline
 \end{tabular}
 \label{tab-specfit}
 \end{minipage}
 \end{table*}

\begin{figure*}
\centering
\begin{tabular}{cc}
\includegraphics[scale=0.48]{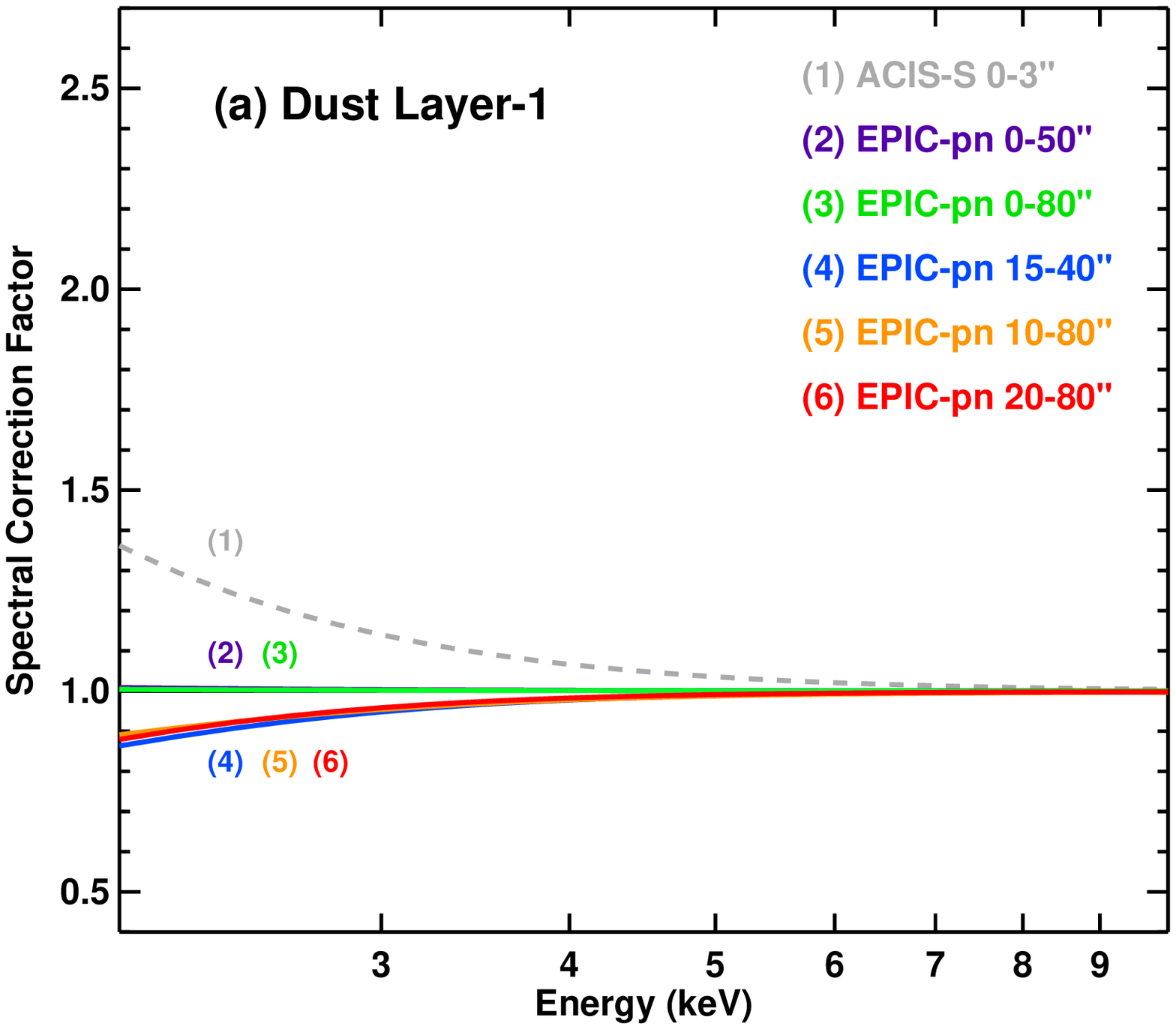} &
\includegraphics[scale=0.48]{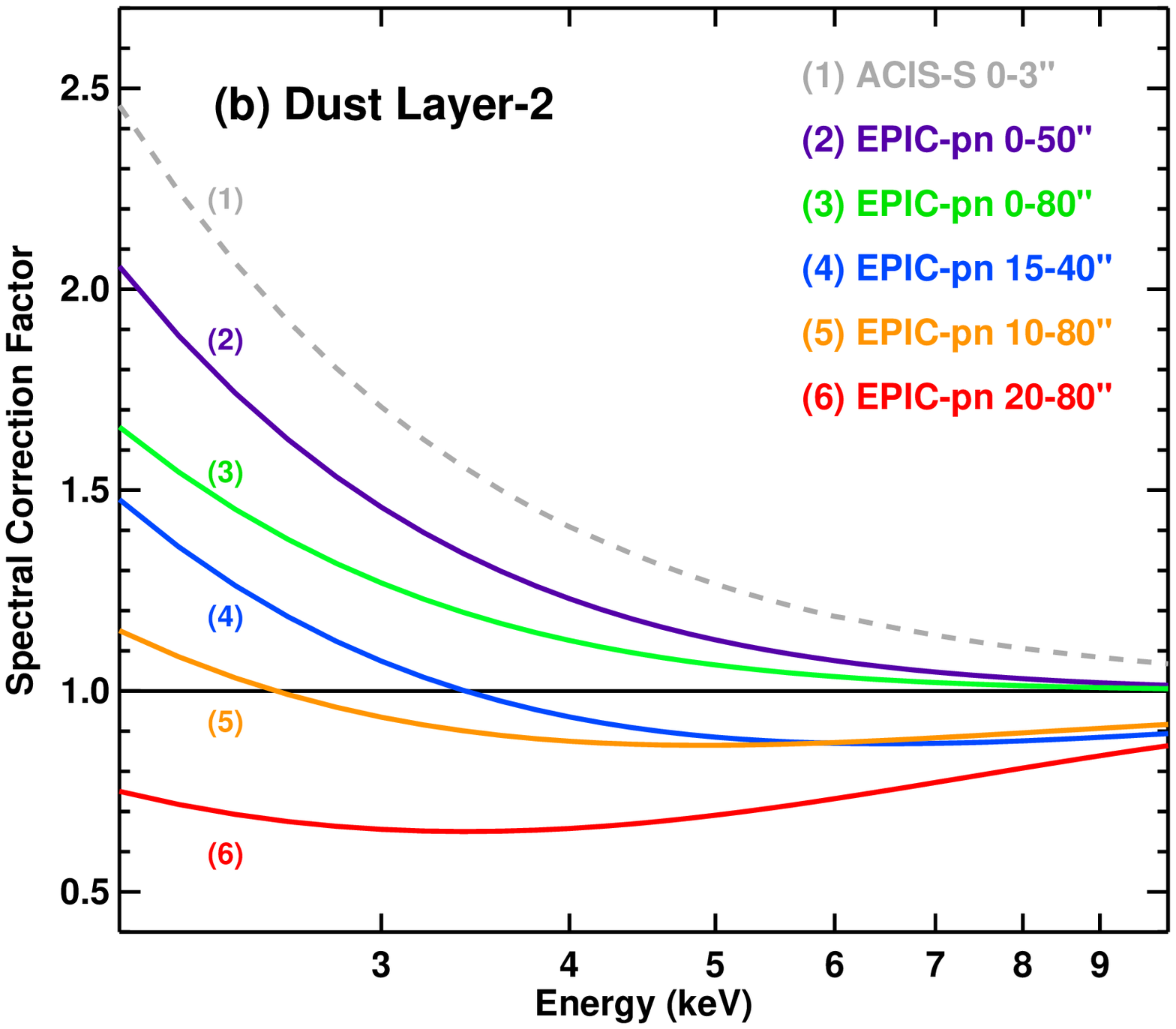} \\
\end{tabular}
\caption{Energy dependent spectral correction factor for the two dust layers for various instruments and spectral extraction regions, based on the best-fit parameters with the COMP-AC-S dust grains (Table~\ref{tab-radfit}). Panel-a is for Layer-1 which is local to \axj; Panel-b is for Layer-2 which is far from \axj. It is clear that Layer-2 introduces much bigger spectral biases than Layer-1.}
\label{fig-specratio2}
\end{figure*}

\subsection{Spectral Correction for the Dust Scattering halo}
\label{sec-dscor}
Based on the best-fit two-layer model, we can perform corrections on the spectral effect of the dust scattering halo. We assume the PSF correction has been performed on the spectra extracted from different annulus regions, so the spectral correction is essentially the correction for the difference between the observed source radial profile and the radial profile of the instrumental PSF, with the same flux in the source extraction region. Therefore, this correction only requires well-constrained radial profiles of the source at various energies, but does not require the knowledge about intrinsic properties of the intervening dust.

We defined the spectral correction factor as $F_{int,E}/F_{obs,E}$, where $F_{obs,E}$ and $F_{int,E}$ are the flux at energy $E$ before and after the dust scattering correction, and used the best-fit model with the COMP-AC-S dust grain to calculate the correction factor. Fig.\ref{fig-specratio1} shows the correction factor as a function of energy for different source extraction regions. A bigger correction factor is found at lower energies with smaller extraction regions, because in this case a bigger halo flux loss is expected. Therefore, for instruments such as \cxo\ ACIS where a small source extraction regions is often used, it is more important to consider the spectral effect of the dust scattering halo (Corrales et al. 2016; Smith, Valencic \& Corrales 2016).

Based on the spectral correction factor in Fig.\ref{fig-specratio1}, we can correct the observed spectra for the dust scattering halo and produce a `dust free' spectrum. Fig.\ref{fig-specdif} shows that the two spectra from different source extraction regions become consistent with each other after the spectral correction. However, this correction to the spectral file itself is only approximate as it does not incorporate the full response profile of every energy channel in EPIC-pn, and so the most appropriate way is to correct the model rather than the spectral file. Therefore, we built an {\sc Xspec} multiplicative model ({\sc axjdust}) for the dust scattering halo around \axj\ for different annulus regions and instruments\footnote{{\sc axjdust} model is available upon request.}. We found that adding this model component can fully account for the spectral discrepancy in Fig.\ref{fig-specdif} and produce consistent best-fit parameters from the two apparently different spectra. The best-fit parameters also changes significantly after the inclusion of {\sc axjdust} model (see Table~\ref{tab-specfit}). Since the absorbed power law model is clearly not good enough for the entire 2-10 keV band, we restrict the fitting to 3-6 keV band and we find a much better fit with $\chi^2/dof = 595/596$, $N_{H,abs} = (3.02\pm0.17)\times10^{23} cm^{-2}$ and $\Gamma = 2.02\pm0.17$. These parameters are also roughly consistent with those reported by Paizis et al. (2015) using the lower S/N \cxo\ HETG spectrum and applying the same absorbed power law model with the same cross-sections and abundances.

Furthermore, to assess the contribution from different dust layers, we calculated the spectral correction factor assuming there is only dust layer-1 or layer-2, separately. It is found that the bias is mainly caused by dust layer-2 which is far from \axj\ (Fig.\ref{fig-specratio2}). This is because dust layer-1 is so close to \axj\ that its halo is very compact and its shape very similar to the instrumental PSF (see Fig.\ref{fig-radfit}), and so the PSF correction also corrects for the halo flux loss with small biases. Dust layer-2 is far from \axj, and so it produces an extended dust scattering halo, which is more extended than the instrumental PSF and is also more difficult to be covered by a small spectral extraction region, thereby requiring a more significant correction factor.

\begin{figure*}
\centering
\begin{tabular}{cc}
\includegraphics[bb=72 144 648 504, scale=0.45]{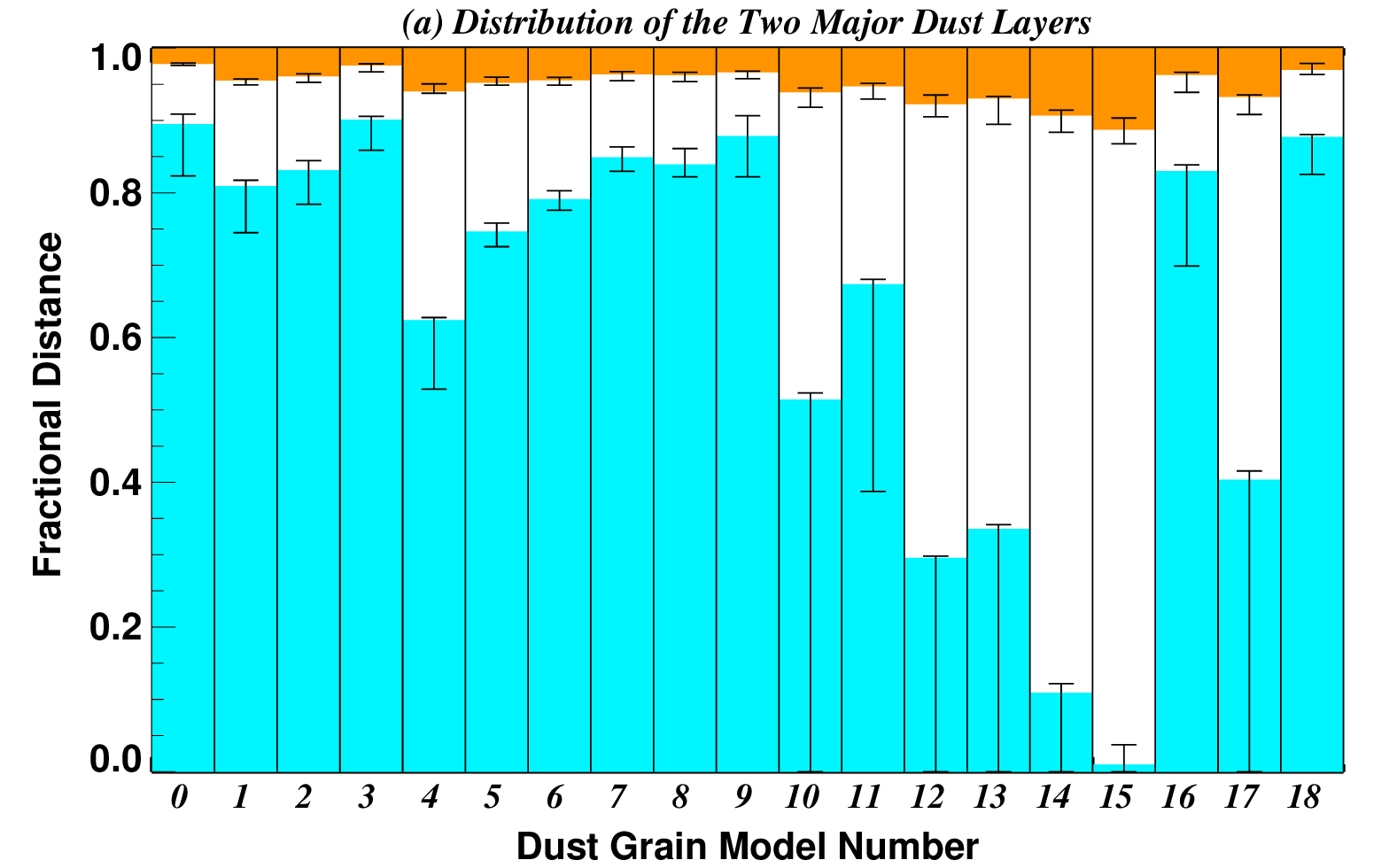} &
\includegraphics[bb=72 144 432 468, scale=0.5]{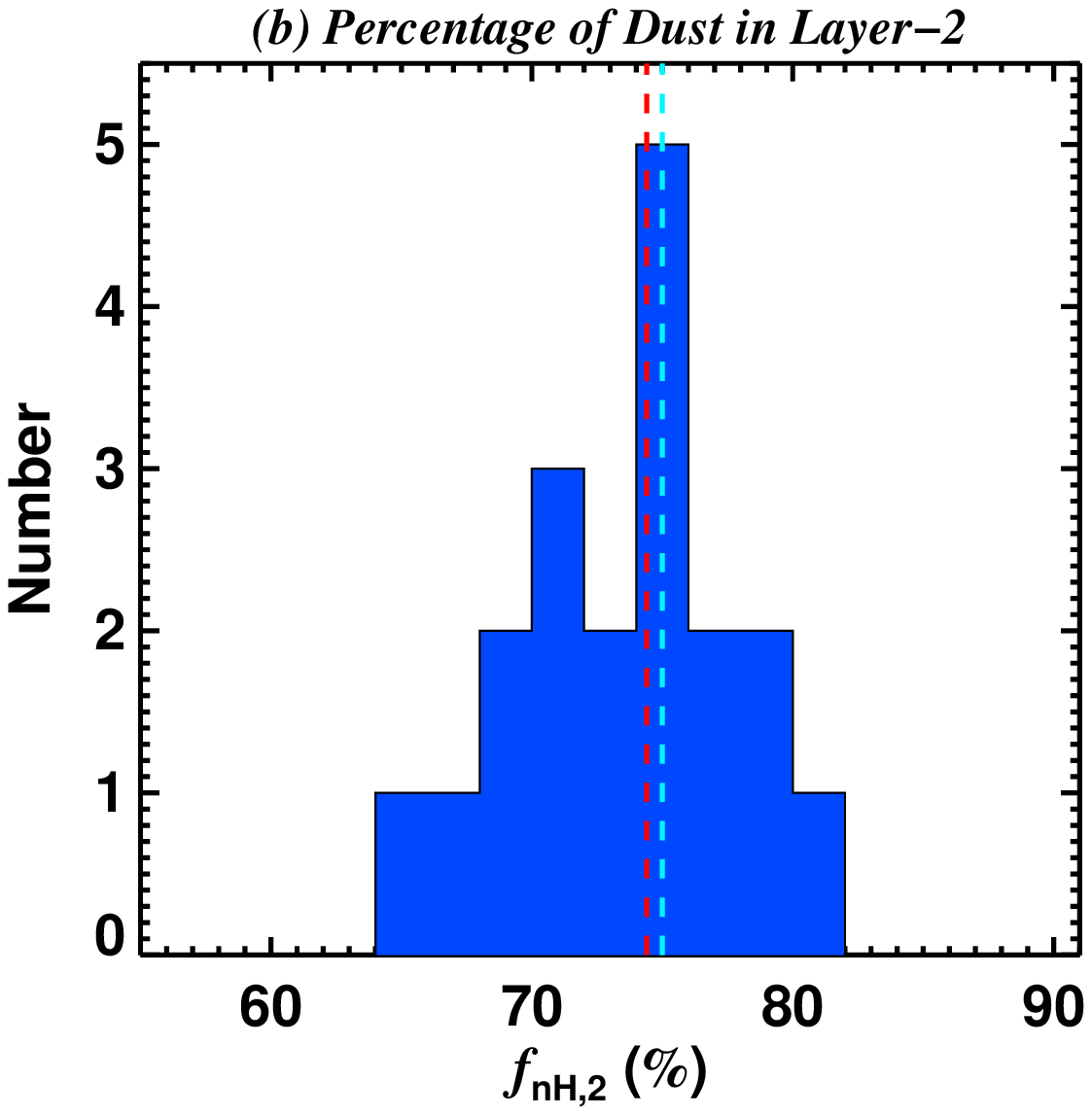} \\
\end{tabular}
\caption{Panel-a: the distribution of the two major dust layers along the LOS, with the orange region being Layer-1 and cyan region being Layer-2. Distance 0 corresponds to the Earth, while distance 1 corresponds to \axj. Error bars indicate the statistical uncertainty of each layer's boundary. Significant separation between the two layers can be found in very dust grain model. Panel-b: the distribution of the percentage of LOS dust in dust Layer-2. The red and cyan dash lines indicate the mean value of 74.4\% and median value of 75.0\%, separately. Dust grain model number and all the data can be found in Table \ref{tab-radfit}, \ref{tab-tau}.}
\label{fig-hist}
\end{figure*}

\section{Discussion}

\subsection{Dust Distribution in the GC Direction}
\label{sec-distribution}
Interstellar dust is typically contained in molecular clouds, whose spatial distribution can be traced by molecular lines (e.g. CO emission, Dame et al. 2001). Previous Galactic CO maps showed a prominent feature due to a massive molecular ring at 4-5 kpc from the GC (the so-called 5 kpc molecular ring, Jackson et al. 2006; Simon et al. 2006), or 3-4 kpc from Earth in the GC direction (Valle\'{e} 2014). Later studies showed an enhancement of mass density of the molecular clouds between 3-6 kpc from the GC overlapping with the 5 kpc molecular ring, which also seems to be distributed along the Scutum Arm ( Roman-Duval et al. 2010; Dobbs \& Burkert 2012; Sato et al. 2014; Heyer \& Dame 2015). Previous surveys about the infrared dark clouds (Egan et al. 1998; Carey et al. 1998) found these clouds mainly lie along the Scutum arm between 3-6 kpc from the GC (Simon et al. 2006; Marshall, Joncas \& Jones 2009) with typical $N_{H}\ga10^{22}cm^{-2}$ in a single cloud (Simon et al. 2006; Peretto \& Fuller 2010). Therefore, it is indeed possible for the GC LOS to contain a large amount of intervening dust about half way between the Earth and the GC.

Another method to infer the dust distribution is to use the dust extinction map. Using the infrared extinction curve towards the GC, Fritz et al. (2011) reported that most of the extinction is caused by the dust in the Galactic disk rather than in the Nuclear bulge (Mezger et al. 1996). Schultheis et al. (2014) built a 3D dust extinction map for the Galactic bulge and found a peak of extinction at 3 kpc from the GC, implying a dust lane in front of the Galactic bar, but they cannot rule out the possibility that this peak is associated with the 5 kpc molecular ring.

All the above studies consistently show that most of the dust is distributed several kpc from the GC in the Galactic disk. This invalidates the assumption made in Tan \& Draine (2004), where the dust scattering halo from \sgra\ was assumed to rise from the intervening dust and gas close to \sgra. Interestingly, recent studies of PSR J1745-2900, the magnetar located at only 2.4 arcsec from \sgra\ (Kennea et al. 2013; Mori et al. 2013; Rea et al. 2013), also show that most of the scattering medium responsible for the broadening of the radio emission from this magnetar is likely associated with the interstellar gas in the Galactic disk, possibly in a nearby spiral arm, rather than close to the GC (Eatough et al. 2013; Shannon \& Johnston 2013; Bower et al. 2014; Wucknitz 2015; Sicheneder \& Dexter 2016; but see Spitler et al. 2014).

Our study on \axj\ is the first detailed observation and modelling of a dust scattering halo around a source in the GC. The dust properties and distribution of the two major layers show a large dispersion, which is mainly due to the lack of knowledge about the properties of dust grains in the ISM especially along the GC LOS. But there are some robust results that can be derived from all the dust grain models. Firstly, although the separation between the two major layers changes from 0.09 to 0.88 (with mean value: 0.31) for different dust grain models, all the models consistently find a highly significant separation between the two layers (see Fig.\ref{fig-hist}a), i.e. Layer-1 is always found to be local to \axj, while Layer-2 is significantly far away and always has its lower boundary reaching 0. Secondly, we find that the percentage of dust that is contained in each layer is tightly constrained and it is observed to be (66-81)\% (with mean value: 74.4\%) in Layer-2, with a much weaker dependence on the dust grain models than the other parameters (see Fig.\ref{fig-hist}b). We emphasize that an extra uncertainty of this dust distribution lies in the assumption of the same type of dust grains in different dust layers. It has been reported that the abundances in the GC may be higher than in the Galactic disk (see Section~\ref{sec-dgratio} for more detailed discussion), and the GC dust grains might have a smaller characteristic size (Hankins et al. 2017). However, there are no strong constraints on these GC parameters, and we are not sure if \axj\ is located inside the GC region or not (see discussions below), thus in this work we do not consider different dust grain models for different dust layers.

\begin{figure}
\centering
\includegraphics[bb=155 0 612 700, scale=0.29]{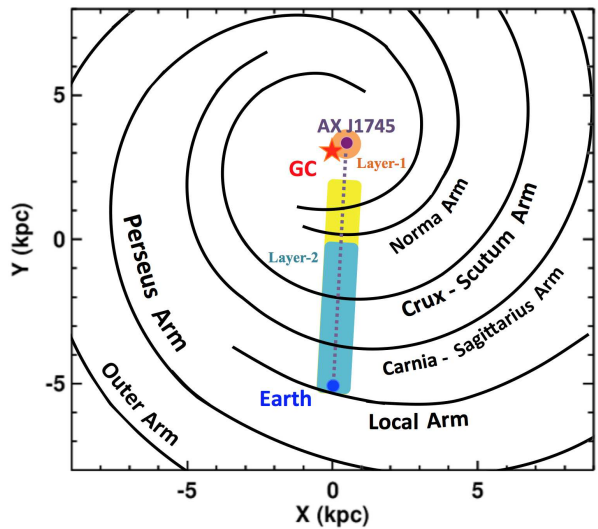}
\caption{Schematic of the Milky Way spiral arms viewed from the north Galactic pole based on the Milky Way structure reported by Nakanishi \& Sofue (2016), Reid et al. (2009) and Caswell \& Haynes (1987). The orange and dark green regions indicate the location of dust layer-1 and dust layer-2 based on the average best-fit parameters in Table~\ref{tab-radfit}, assuming \axj\ is in the GC. The yellow region shows the highest upper boundary (0.9) of dust layer-2 from the fitting using the BARE-GR-B dust grain model.}
\label{fig-gcstruct}
\end{figure}

The absolute location of these two layers also depends on the absolute distance of \axj\ itself, which is still highly uncertain. From the fact that the $N_{H,abs}$ of \axj\ is significantly larger than that of \sgra\ (fusing the same absorption model and abundances, see Section~\ref{sec-axj}), it can be reasonably inferred that \axj\ is NOT closer to the Earth than \sgra. Although $N_{H,sca}$ is not directly comparable to the $N_{H,abs}$ as they have different normalizations (see Section~\ref{sec-dgratio}), by assuming the same dust properties in all the layers, we can use the fraction of dust to infer the fraction of $N_{H,abs}$ in each layer. Therefore, adopting $N_{H,abs}=3\times10^{23}cm^{-2}$ for \axj\ (see Section~\ref{sec-dscor}) and $1.6\times10^{23}cm^{-2}$ for \sgra\ from the {\sc tbnew} model and WAM00 abundances (Ponti et al. 2017b), we can estimate $N_{H,abs}=(2.0-2.4)\times10^{23}cm^{-2}$ in Layer-2, which is similar to \sgra. This is also consistent with the fact that Layer-1 is local to \axj\ and Layer-2 is much farther away, thus it is likely that only Layer-2 is in front of \sgra.

If a distance of 8 kpc is adopted for \axj, the mean fractional distance of Layer-2 would indicate that it is is 0-5 kpc from the Earth in the Galactic disk, thus overlaps with the Crux-Scutum Arm, Carnia-Sagittarius Arm and the local Arm (see Fig.\ref{fig-gcstruct}; Nakanishi \& Sofue 2016; Reid et al. 2009; Caswell \& Haynes 1987). This result immediately implies the possibility that the same dust in Layer-2 may also intervene along the LOS of many GC sources. There are two X-ray transients whose LOSs are very close to \sgra. Swift J174540.7-290015 was discovered in 2016 at 16 arcsec from \sgra\ (Reynolds et al. 2016), with $N_{H,abs}=1.7\times10^{23}cm^{-2}$ ({\sc tbnew} model and WAM00 abundances, Ponti et al. 2016). SGR J1745-2900 is located at 2.4 arcsec from \sgra\ (Rea et al. 2013), with $N_{H,abs}=1.9\times10^{23}cm^{-2}$ ({\sc tbabs} model and WAM00 abundances, Coti-Zelati et al. 2015). Although a further correction for the dust scattering opacity is probably also necessary for these GC sources, their $N_{H,abs}$ are indeed similar to that of Layer-2 (also see Ponti et al. 2017b), which support Layer-2 being a common GC-foreground dust layer in the Galactic disk. A more conclusive study would be to compare the dust scattering halo around many GC sources and see if a similar dust component as Layer-2 exists in their halo profiles.

\subsection{Testing Dust-to-Gas Relations in the GC Direction}
\label{sec-dgratio}
Since $\tau_{sca}$, $N_{H,sca}$ and $N_{H,abs}$ have all been measured independently in \axj, it becomes the first source that allows the test of dust-to-gas relations in the GC LOS. However, we would like to emphasize that there are some underlying systematic uncertainties associated with the $N_{H}$ measurement. Firstly, $N_{H,sca}$ depends heavily on the assumptions made in the dust grain models such as the abundances and the grain size distribution, which affect the halo intensity and profile, which then affect $N_{H,sca}$ during the halo profile fitting. This is why a large dispersion of $N_{H,sca}$ is found among all the dust grain models. Secondly, $N_{H,abs}$ also depends heavily on the assumed abundances during the X-ray spectral fitting. In the case of \axj, we found $\sim$50\% change in $N_{H,abs}$ by simply changing the abundances from the AG89 solar abundances to WAM00 ISM abundances. Moreover, $N_{H,abs}$ depends on the absorption model and intrinsic spectral model used for the spectral fitting. A high S/N and resolution spectra with a good intrinsic model would allow the absorption model to determine $N_{H,abs}$ from various metal absorption edges; while a low S/N spectra or a bad intrinsic model would force the absorption model to trace the spectral curvature (as likely in \axj\ and many other GC sources where the strong extinction leads to a fast decrease of S/N in the soft X-ray band), which would bias $N_{H,abs}$. Therefore, it is not meaningful to do a precise comparison between $N_{H,sca}$ and $N_{H,abs}$. Actually, some of the dust and gas in Layer-1 can be so close to \axj\ that they would absorb X-rays but not {\it observably} scatter them, and so the {\it observed} $N_{H,sca}$ should be smaller than the intrinsic $N_{H}$. It must also be stressed that it is necessary to provide key informations when an $N_{H,abs}$ is reported, such as the quality of spectra being fitted, models used for the intrinsic spectra and X-ray absorption, and assumptions made for the cross-sections and abundances.

On the other hand, several versions of the dust-to-gas relations between $N_{H,abs}$, $A_{V}$ and $\tau_{sca}$ have been reported so far. For example, Predehl \& Schmitt (1995) (PS95) performed X-ray spectral and halo fittings for 25 point sources observed by {\it ROSAT} and reported $N_{H,abs}/A_{V}=1.79\times10^{21}cm^{-2}mag^{-1}$ and $\tau_{sca}=0.087\times A_{V}(mag)~E(keV)^{-2}$. Their absorption model used the Morrison \& McCammon (1983) cross-sections, which is equivalent to the {\sc Xspec} {\sc wabs} model. However, Draine \& Bone (2004) (DB04) argued that Predehl \& Schmitt (1995) may have underestimated the scattering power and so a better relation would be $\tau_{sca}=0.15\times A_{V}(mag)~E(keV)^{-1.8}$. Later G\"{u}ver \& \"{O}zel (2009) (LO09) collected 22 supernova remnants which have their $N_{H,abs}$ measured from the high-quality X-ray spectra from \cxo, \xmm\ and {\it Suzaku}, and have independent $A_V$ measurements. They reported $N_{H,abs}/A_{V}=(2.21\pm0.09)~\times10^{21}cm^{-2}mag^{-1}$. Their difference from Predehl \& Schmitt (1995) was attributed to the narrow bandpass of {\it ROSAT} and the simple power law model assumed for the intrinsic spectra. However, it is difficult to track down all the absorption models, cross-sections and abundances used to derive the $N_{H,abs}$ of their sample because they were reported by many different papers and authors. Recently, Valencic \& Smith (2015) (VS15) analyzed the X-ray dust scattering halo around 35 sources and found $N_{H,abs}/A_{V}=(2.08\pm0.26)~\times10^{21}cm^{-2}mag^{-1}$, with $\tau_{sca}(1.0keV)/A_{V}$ having a typical $\pm1\sigma$ range of 0.02-0.08 depending on the dust grain model. Their $N_{H,abs}$ was based on the AG89 solar abundances and {\sc phabs} model.

In order to use these relations, we adopt $N_{H,abs}\simeq2\times10^{23}cm^{-2}$ for \axj\ also based on the {\sc phabs} model and the AG89 solar abundances. It is known that the metallicity near/in the GC is higher than the solar abundances (e.g. Kubryk, Prantzos \& Athanassoula 2015 and references therein). If some of the dust and gas in e.g. Layer-1 is indeed close to the GC, the $N_{H,abs}$ from the AG89 abundances would have been over-predicted. However, it is also reported that the AG89 metallicity over-estimated the solar metallicity (e.g. Asplund et al. 2009; Nieva \& Przybilla 2012), thus the adoption of AG89 abundances may actually alleviate the potential higher-metallicity problem towards \axj. We find $A_{V}=117,~90,~96~mag$ from the $N_{H,abs}/A_{V}$ relations in PS95, LO09 and VS15, respectively. Adopting $A_{V}\simeq100~mag$, we derive $\tau_{sca}=$ 0.80, 0.35 and 0.18 at 3.3, 5.0 and 7.0 keV from the $\tau_{sca}/A_{V}$ relation in PS95, or $\tau_{sca}=$ 1.75, 0.83 and 0.45 from DB04, or $\tau_{sca}=$ 0.73, 0.32 and 0.16 from VS15. Therefore, it appears that the dust-to-gas relations in PS95 and VS15 give more consistent $\tau_{sca}$ with those measured directly from the eclipsing light curves. 

\begin{figure}
\centering
\includegraphics[scale=0.45]{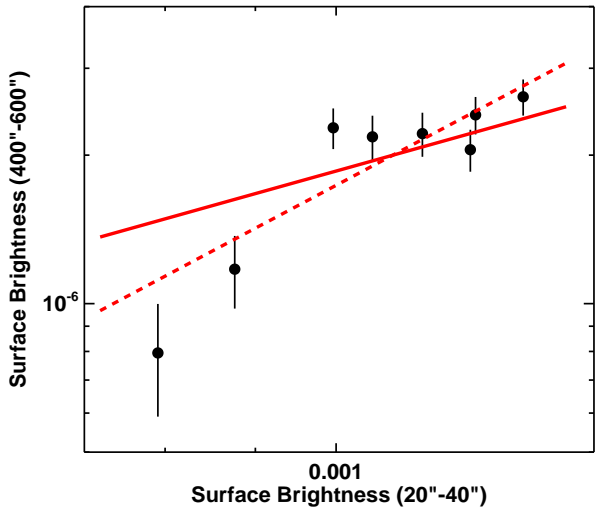}
\caption{The correlation between the halo wing and halo core intensity in the 4-6 keV band. The unit for the surface brightness is $counts~cm^{-2}s^{-1}arcmin^{-2}$. The data points are measured from the eight \xmm\ observations in Table~\ref{tab-obs}. The red solid line is the best-fit line assuming a slope of 1, the red dashed line is the best-fit line with the best-fit slope of $1.9\pm0.5$.}
\label{fig-wingratio1}
\end{figure}

\subsection{Excess Flux in the Halo Wing}
\label{sec-wing}
In this work, we reported, for the first time, the detection of a significant wing of the dust scattering halo at $E\ge5~keV$.
On the other hand, previous studies about the dust scattering halo mainly focussed on energies below 3 keV (e.g. Predehl \& Schmitt 1995; Xiang, Zhang \& Yao 2005; Valencic \& Smith 2015), and used a free constant to fit the underlying background, hampering the distinguishing of an extended halo wing even if it existed. This is possible in our work because the background is well constrained from observations where \axj\ was in quiescence, which then enabled us to detect this excess flux in the halo wing.

In order to verify that this halo wing is related to the scattering light from \axj, we tested the correlation between the flux of the halo wing and the flux of the halo core using all the \xmm\ observations in Table~\ref{tab-obs}. Fig.\ref{fig-wingratio1} shows that a clear correlation is found. The Spearman's rank correlation coefficient is 0.79 and the $p$-value is 0.02, and Pearson's linear correlation coefficient is 0.89. A linear fitting with slope fixed at 1 would produce a $\chi^2$ of 7.1 for 7 dof (red solid line in Fig.\ref{fig-wingratio1}). Freeing the slope in the fitting only improves the $\chi^2$ by 1.0 for 1 extra free parameter, and the best-fit slope is $1.9\pm0.5$ (red dashed line in Fig.\ref{fig-wingratio1}). These results indicate the correlation between the halo wing flux and halo core flux is consistent with a linear correlation, confirming that the wing structure in the radial profile is indeed part of the dust scattering halo around \axj.

Since smaller dust grains can scatter photons at larger angles more efficiently, this halo wing implies an excess fraction of dust grains of relatively small sizes in the ISM. According to Eq.\ref{equ-awing}, for a viewing angle of 300-600 arcsec at 5 keV, the typical dust grain size would be 300-590 \AA, which is still two orders of magnitude bigger than the wavelength of X-ray photons (2.5 \AA\ for 5 keV), and so the Rayleigh-Gans approximation is still valid. However, the outer boundary of the halo wing is still unknown as at larger radii the wing starts to blend with the diffuse and background emission. If the wing is more extended than currently detected, the inferred dust grains can have smaller sizes. Therefore, the halo wing suggests a higher fraction of dust grains with typical sizes of $\lesssim$590 \AA\ than considered in current dust grain models. This size range covers the typical sizes of PAH (3.5-55 \AA, ZDA04), and is within the typical sizes of Graphite, ACH2, Olivine, Enstatite, Fe metal and the composite dust grains in ZDA04 (but is one order of magnitude smaller than the largest dust grains). Since we do not know the size distribution and composition of this dust, it is not possible to predict the influence in other aspects such as the IR emission or excitation. Future studies of the dust grain model along the GC LOS should take these halo wing phenomena into account.

\subsection{Halo Uniformity}
\begin{figure}
\centering
\includegraphics[bb=0 0 612 650 ,scale=0.33]{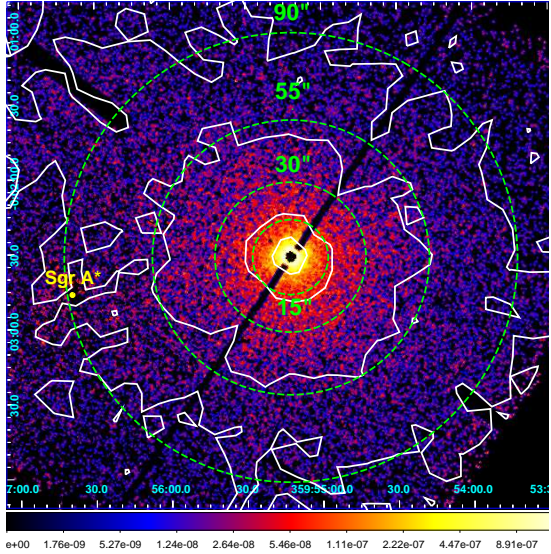}
\caption{Background subtracted image of \axj\ in the 4-6 keV band using ACIS-S observations. The vertical axis is the Galactic latitude, the horizontal axis is the Galactic longitude. The readout streaks and central 2.5 arcsec are subtracted. The white surface brightness contours are based on the same image smoothed by 15 arcsec, which shows no strong azimuthal variation except for the artificial non-uniformity due to the readout streaks.}
\label{fig-asymmetry}
\end{figure}

The non-uniformity of dust scattering halos has been noticed before (McCollough, Smith, Valencic 2013; Seward \& Smith 2013; Valencic \& Smith 2015), which may rise from partially-aligned non-spherical grains (Draine \& Allaf-Akbari 2006) or azimuthal variation of the foreground dust column density. Following our previous discussion, if \axj\ is located inside or beyond the GC, then Layer-1 is probably associated with the molecular clouds in the Central Molecular Zone (CMZ), whose column density is known to vary severely along different LOS (e.g. Serabyn \& Guesten 1987; Morris \& Serabyn 1996; Molinari et al. 2011; Ponti et al. 2013; Langer et al. 2015; Henshaw et al. 2016). If so, the halo produced by Layer-1 might show azimuthal variation. To test this, we generated a contour map on the ACIS-S image of \axj\ in the 4-6 keV band (Fig.\ref{fig-asymmetry}, based on ObsID: 17857). No obvious azimuthal variation is seen in the image except for the artificial asymmetry due to the subtraction of the readout streaks, suggesting that Layer-1 should not have strong column density variation. The halo outside 20 arcsec is dominated by Layer-2, and we found no clear asymmetry out to the LOS of \sgra, either. This is supporting evidence that the same Layer-2 may also intervene along the LOS to \sgra. Further study of the non-uniformity would require more photon counts and more careful azimuthal structure analysis (e.g. Seward \& Smith 2013), which is beyond the scope of this work.

\section{Conclusions}
In this paper, we used a large dataset from \cxo\ and \xmm\ to conduct a detailed analysis of the radial profile of \axj\ whose LOS is $\sim$1.45 arcmin away from \sgra.  Being a bright X-ray source with high column density, this source is surrounded by a strong dust scattering halo across the entire 2-10 keV band.\\
\\
(1) Based on the halo profile modelling, we identified two major scattering components, namely dust Layer-1 and Layer-2.  For all the 19 dust grain models used in this work, and assuming the same dust grain model along the LOS, we found Layer-1 is produced by the dust close to \axj\ within a fractional distance of $\lesssim0.11$ and contains (19-34)\% (mean value: 26\%) of the total LOS dust. The remaining (66-81)\% (mean value: 74\%) LOS dust is contained in Layer-2, distributing from the Earth up to a fractional distance of 0.01-0.90, with a mean value of 0.64 from all dust grain models. These two layers separate from each other significantly by a fractional distance of 0.08-0.88, with a mean value of 0.31. Assuming that \axj\ is located at 8 kpc in the GC, Layer-2 must be distributed in the Galactic disk and probably associated with the molecular clouds distributed along the spiral arms on the LOS, such as the Crux-Scutum Arm, Carnia-Sagittarius Arm and the Local Arm. \\
\\
(2) In addition to the two dust scattering components, we identified an extended component in the halo wing which cannot be explained by any dust grain model used in this work. This halo wing suggests a higher fraction of dust grains with typical sizes of $\lesssim$ 590 \AA\ than considered in current dust grain models.\\
\\
(3) We also investigated the influence of the dust scattering halo on the observed spectra. We found that the observed spectra of \axj\ can strongly depend on the source extraction region, which is due to the partial inclusion of the dust scattering halo. The more extended halo from Layer-2 has a stronger spectral effect. After applying the spectral correction for the dust scattering towards \axj, and found $N_{H,abs} = (3.02\pm0.17)\times10^{23} cm^{-2}$ and $\Gamma = 2.02\pm0.17$ by fitting the {\sc tbnew*powerlaw} model to the 3-6 keV spectrum from \xmm\ EPIC-pn, with cross-sections of Verner et al. (1996) and WAM00 ISM abundances.\\
\\
(4) Since Layer-2 is likely associated with the dust in the Galactic disk several kpc away from \axj, it may also intervene the LOS of other nearby GC sources such as \sgra. We found a rough consistency between the $N_{H, abs}$ of Layer-2, \sgra\ and a couple of nearby transients, so it is possible that Layer-2 also produces dust scattering in these sources. A more conclusive study would be to compare the halo shapes from all these sources. If it is confirmed that the same dust population in Layer-2 also intervenes other nearby LOSs towards the GC, a significant spectral correction for the dust scattering halo would be necessary for many GC sources.\\
\\
(5) We built {\sc Xspec} models to account for the spectral discrepancy caused by the entire dust scattering halo around \axj\ ({\sc axjdust} model), as well as the dust scattering in Layer-2 alone ({\sc fgcdust} model) so that it can be applicable to the other sources in the GC.. These models can properly correct for different spectral discrepancies arising from different source extraction regions\footnote{We have created different versions of these dust scattering models for the CCD quality spectra from \xmm\ EPIC, \cxo\ ACIS, \swift\ XRT and {\it NuSTAR}. These models will be made public available. Interested readers can also contact C. Jin directly for these models.}.

\section*{Acknowledgements}
We thank Elisa Costantini for helpful discussions and comments. CJ acknowledges valuable discussions with Lynne Valencic about the cosmic abundances. The anonymous referee is appreciated for providing insightful comments and suggestions, which allowed us to improve the paper significantly. We would also like to thank the \xmm\ team for providing important instructions on technical and calibration issues. This project is supported by the Bundesministerium f\"{u}r Wirtschaft und Technologie/Deutsches Zentrum f\"{u}r Luft- und Raumfahrt (BMWI/DLR, FKZ 50 OR 1408 and FKZ 50 OR 1604) and the Max Planck Society. This work is based on observations obtained with \xmm, an ESA science mission with instruments and contributions directly funded by ESA Member States and NASA. The scientific results reported in this article are also based on observations made by the \cxo\ X-ray Observatory, as well as data obtained from the \cxo\ Data Archive. This research has made use of software provided by the \cxo\ X-ray Center (CXC) in the application packages CIAO, ChIPS, and Sherpa.

\appendix
\onecolumn
\centering
\section{Combined Radial Profiles of \axj\ and the Background Emission}
\begin{figure*}
\centering
\begin{tabular}{ccc}
\includegraphics[bb=54 216 558 648,scale=0.28]{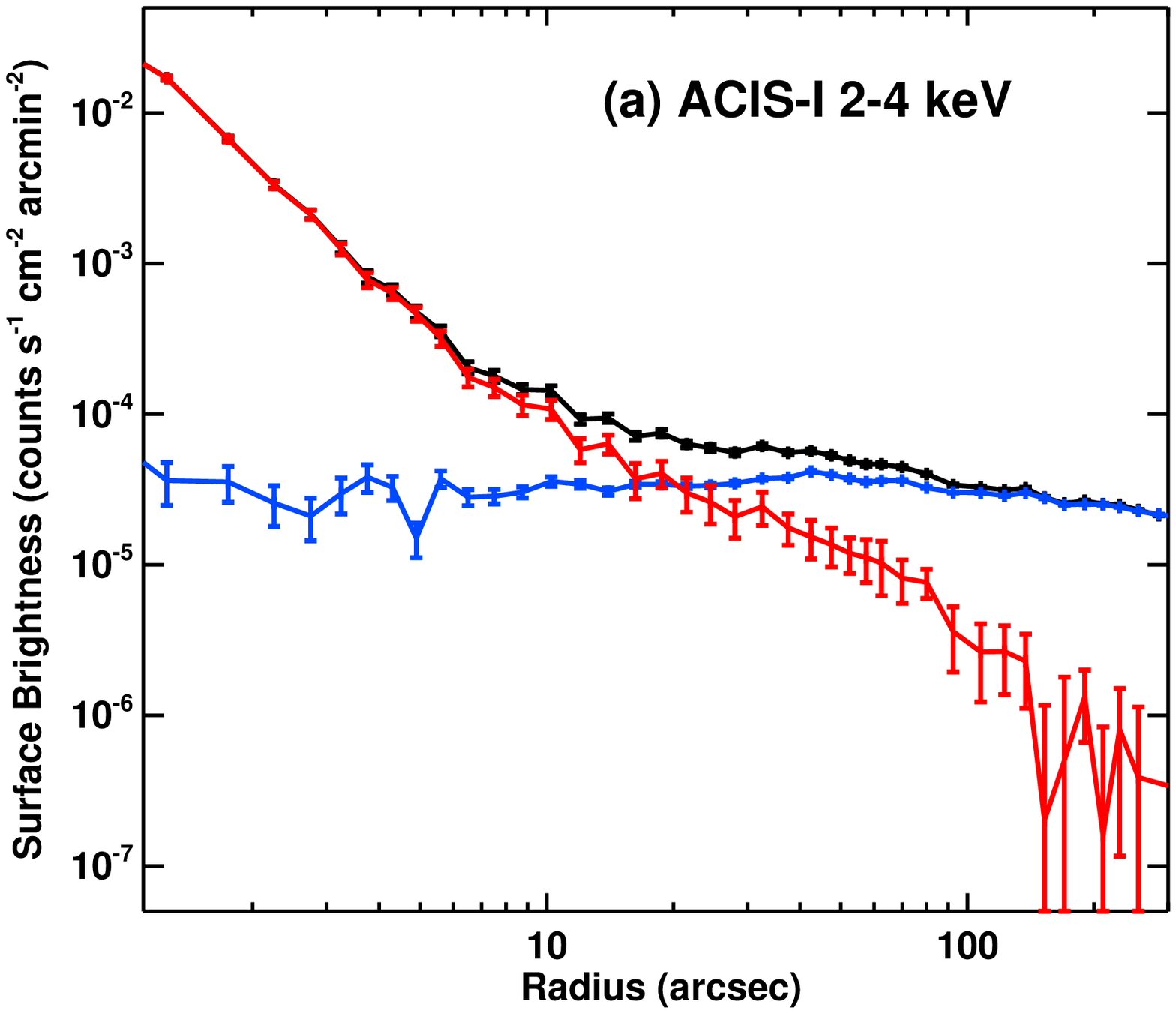} &
\includegraphics[bb=54 216 558 648,scale=0.28]{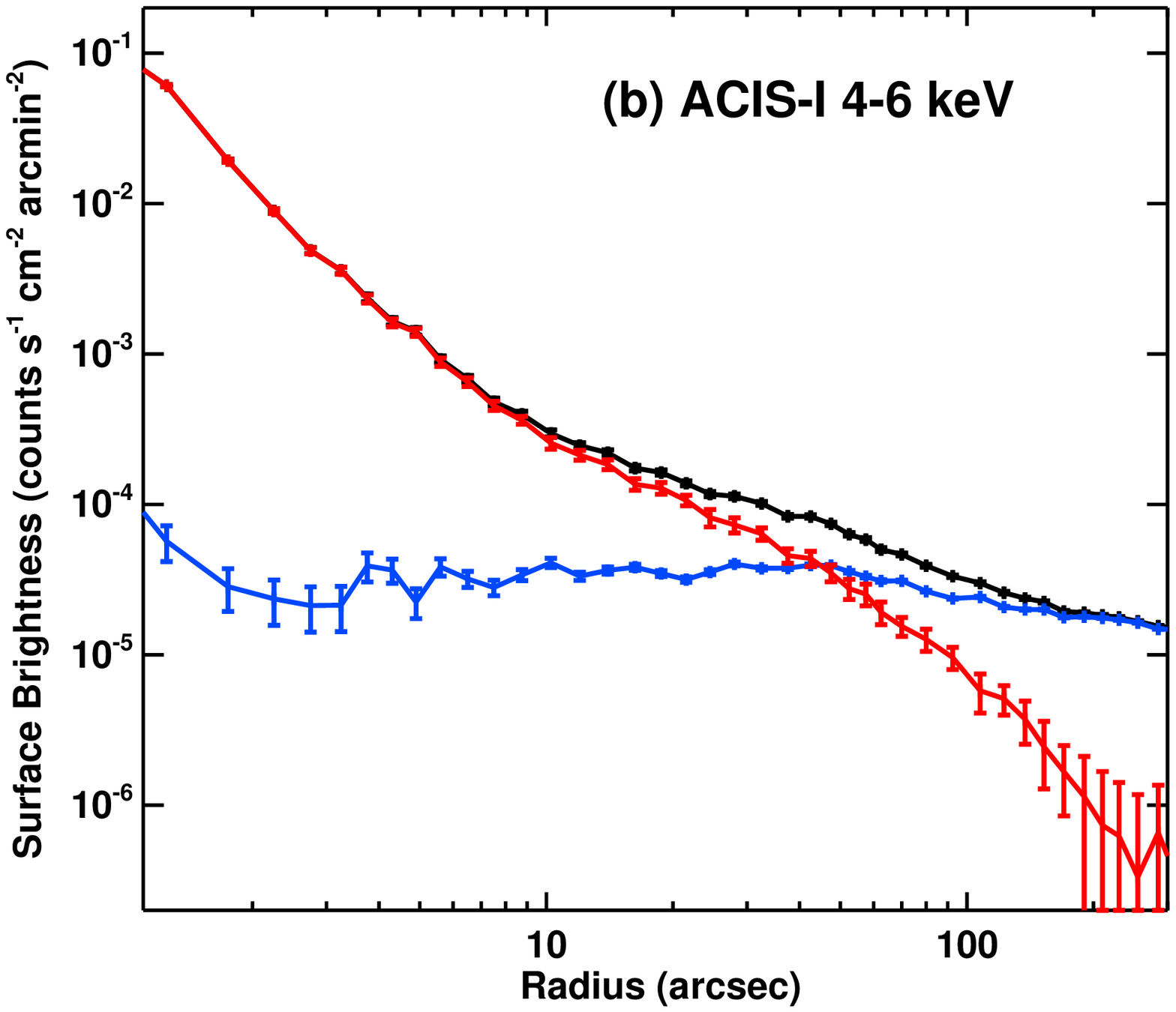} &
\includegraphics[bb=54 216 558 648,scale=0.28]{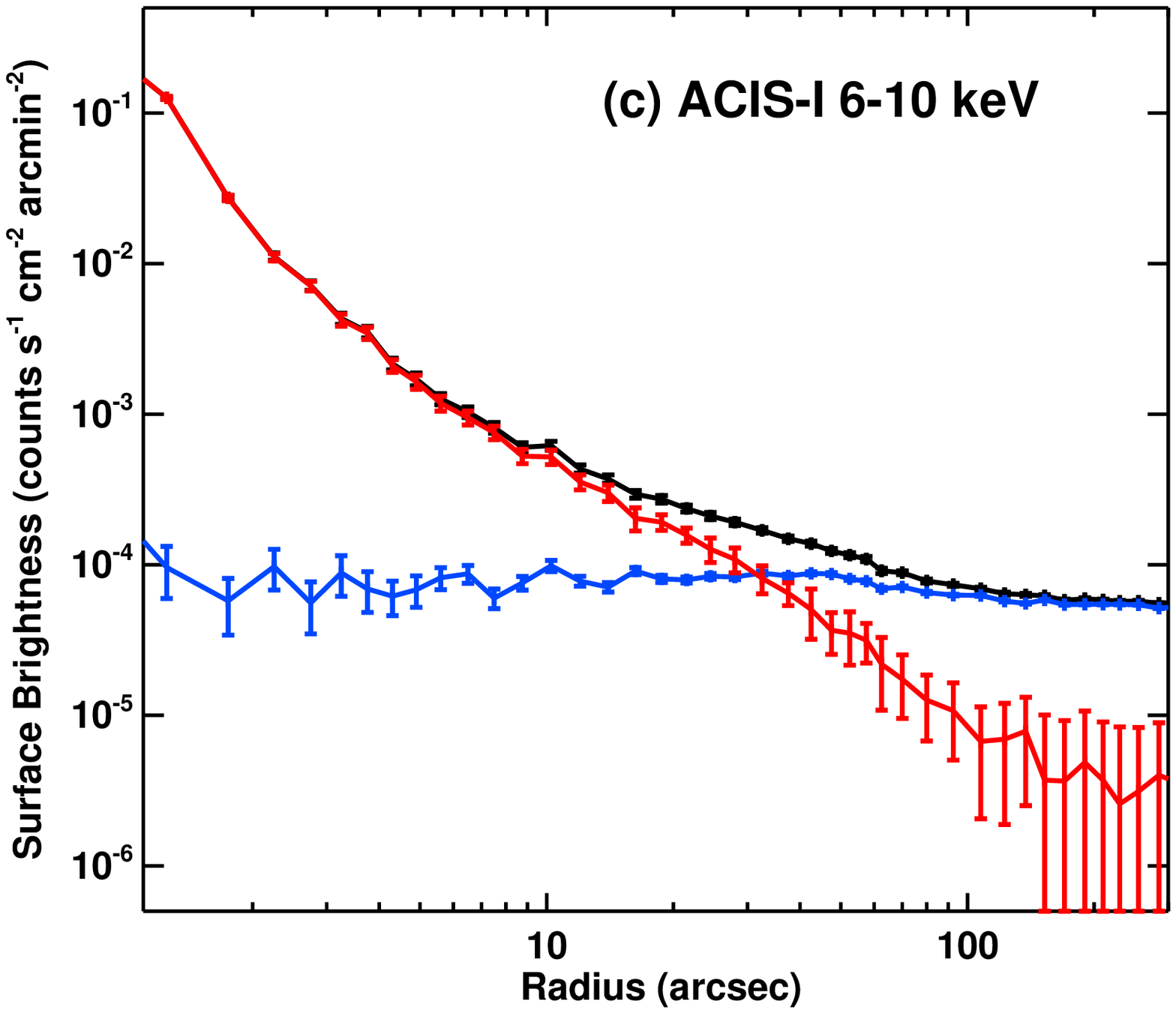} \\
\includegraphics[bb=54 216 558 648,scale=0.28]{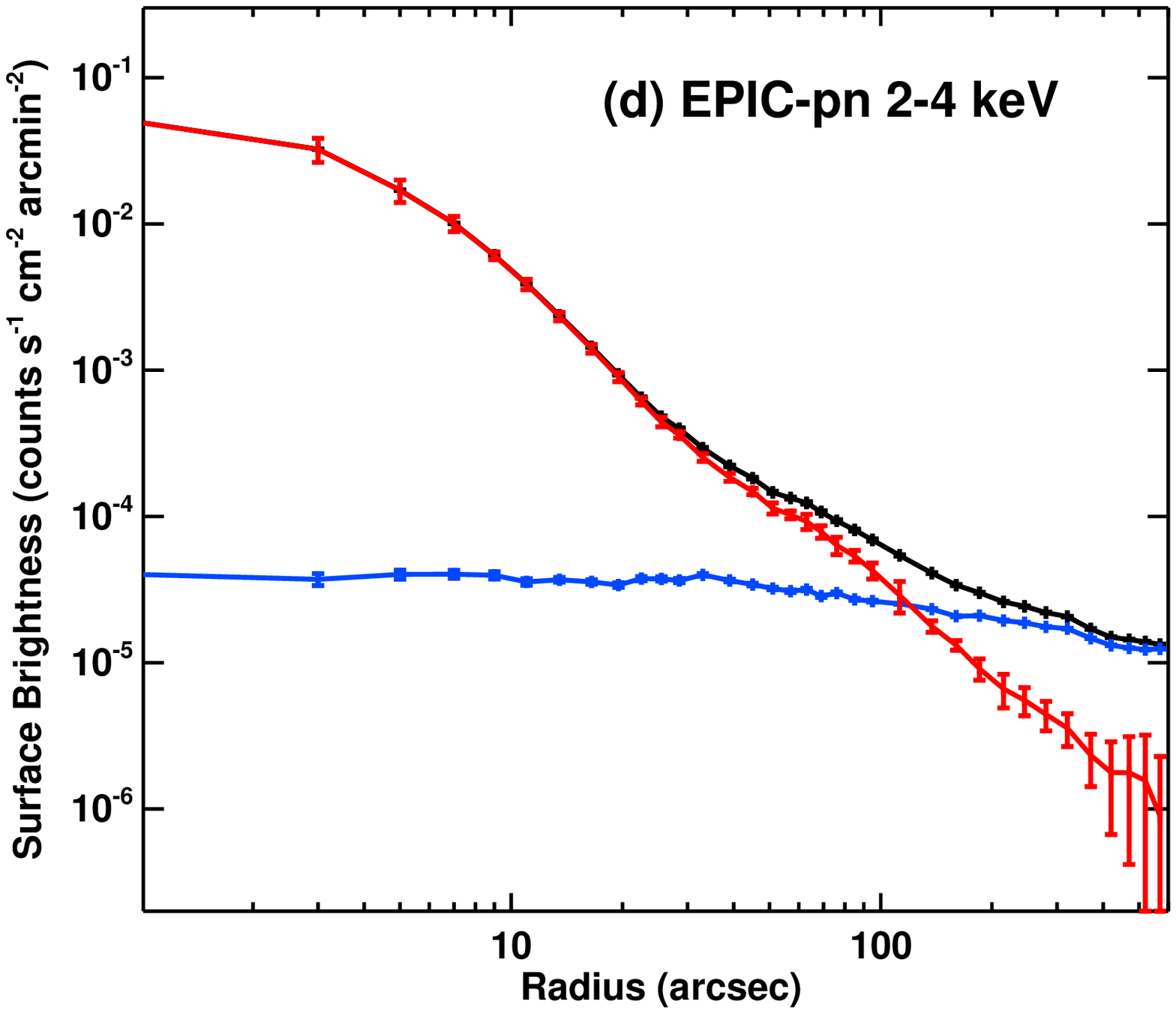} &
\includegraphics[bb=54 216 558 648,scale=0.28]{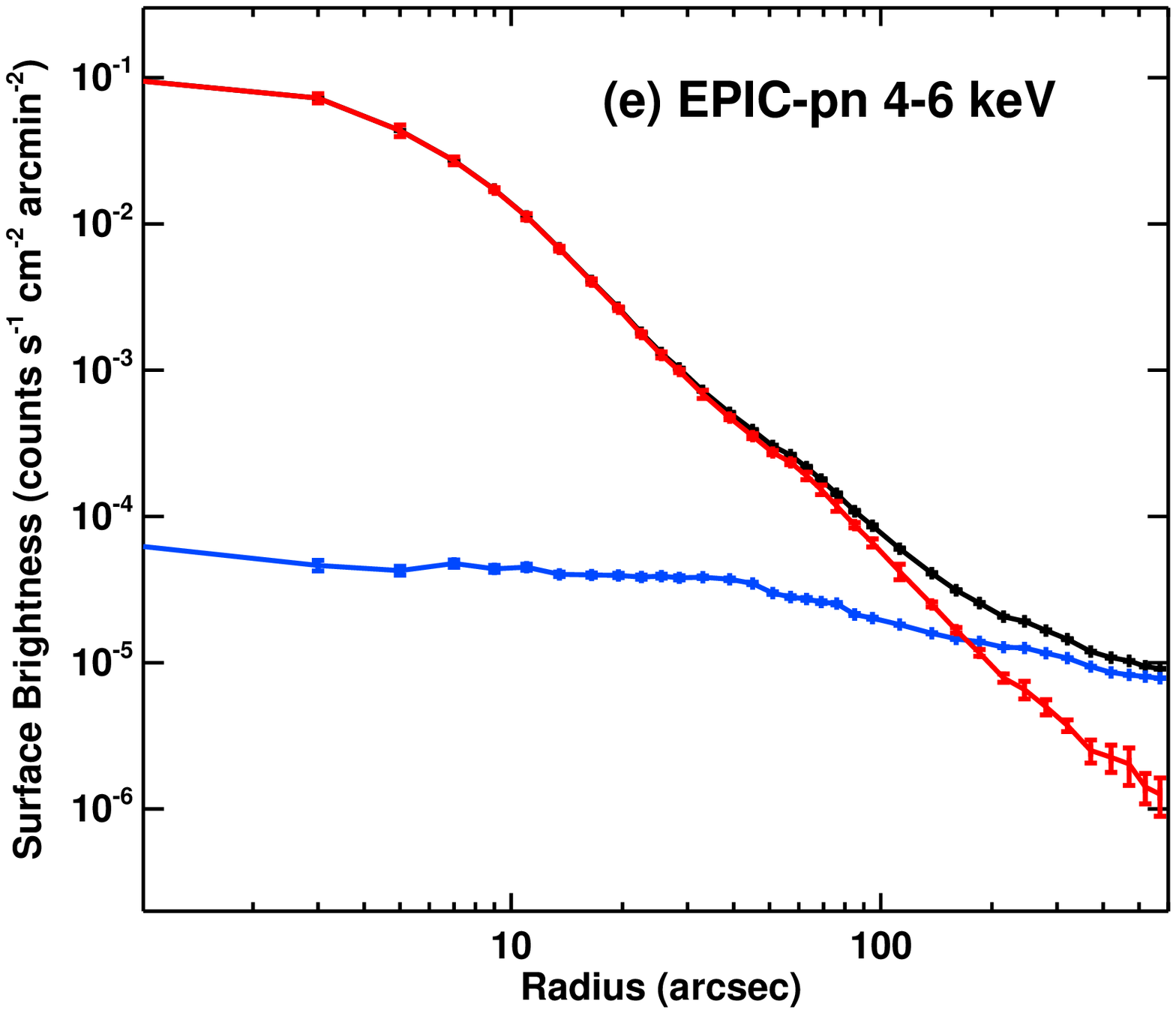} &
\includegraphics[bb=54 216 558 648,scale=0.28]{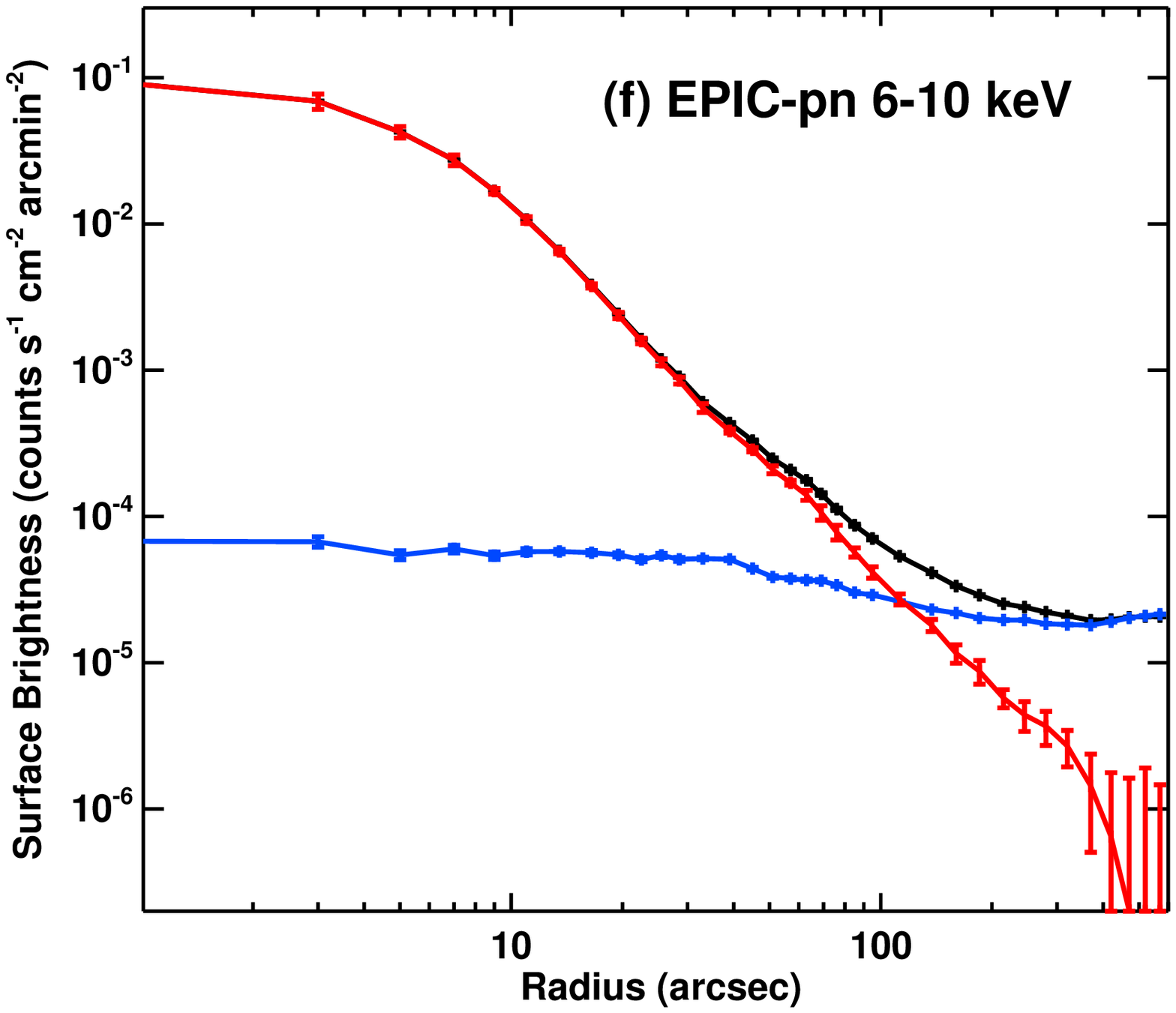} \\
\end{tabular}
\caption{Combined radial profiles of \axj\ in \cxo\ and \xmm\ in three energy bands before background subtraction (black) and after (red). The blue data points show the radial profile of the diffuse emission plus detector background as measured during quiescent periods of \axj.}
\label{app-fig-radprof}
\end{figure*}

\section{List of Background Observations}
\label{app-obslis}
\begin{table*}
 \centering
  \begin{minipage}{170mm}
  \centering
   \caption{List of \xmm\ and \cxo\ observations used to determine the emission underneath the halo of \axj\ when it was in the quiescent state (i.e. detector background plus diffuse emission). `Exp' and `CL-Exp' are the total exposure time and clean exposure time without background flares. $\theta_{off}$ is the off-axis angle of \axj.}
    \label{app-tab-obs}
     \begin{tabular}{ccccccc}
     \hline
     \cxo\ &&&&&&\\
     ObsID & Start-Time & Target & ACIS-Grating & Exp & CL-Exp & $\theta_{off}$ \\
     & & & & (ks) & (ks) & (arcmin) \\
     \hline
     242&1999-09-21 & Sgr A$^\star$&ACIS-I (None)&45.9&42.2&1.14\\
     2943&2002-05-22 & Sgr A$^\star$&ACIS-I (None)&37.7&37.7&1.73\\
     3663&2002-05-24 & Sgr A$^\star$&ACIS-I (None)&38.0&38.0&1.73\\
     4683&2004-07-05 & Sgr A$^\star$&ACIS-I (None)&49.5&49.5&1.23\\
     4684&2004-07-06 & Sgr A$^\star$&ACIS-I (None)&49.5&49.5&1.23\\
     5950&2005-07-24 & Sgr A$^\star$&ACIS-I (None)&48.5&48.5&1.21\\
     5951&2005-07-27 & Sgr A$^\star$&ACIS-I (None)&44.6&44.6&1.21\\
     5952&2005-07-29 & Sgr A$^\star$&ACIS-I (None)&45.3&43.5&1.20\\
     5953&2005-07-30 & Sgr A$^\star$&ACIS-I (None)&45.4&45.4&1.20\\
     5954&2005-08-01 & Sgr A$^\star$&ACIS-I (None)&17.9&17.9&1.20\\
     6363&2006-07-17 & Sgr A$^\star$&ACIS-I (None)&29.8&29.8&1.21\\
     10556&2009-05-18 & Sgr A$^\star$&ACIS-I (None)&112.5&112.5&1.82\\
     11843&2010-05-13 & Sgr A$^\star$&ACIS-I (None)&78.9&78.9&1.81\\
     13016&2011-03-29 & Sgr A$^\star$&ACIS-I (None)&17.8&17.8&1.77\\
     13017&2011-03-31 & Sgr A$^\star$&ACIS-I (None)&17.8&17.8&1.77\\
     14941&2013-04-06 & Sgr A$^\star$&ACIS-I (None)&19.8&19.8&1.78\\
     14942&2013-04-14 & Sgr A$^\star$&ACIS-I (None)&19.8&19.8&1.78\\
     combined & -- & -- & -- & 718.8 & 713.3& -- \\
     \hline
     13850&2012-02-10 & Sgr A$^\star$&ACIS-S (HETG)&59.3&59.0&1.43\\
     14392&2012-02-11 & Sgr A$^\star$&ACIS-S (HETG)&58.4&58.0&1.43\\
     14394&2012-02-10 & Sgr A$^\star$&ACIS-S (HETG)&17.8&17.8&1.43\\
     14393&2012-02-11 & Sgr A$^\star$&ACIS-S (HETG)&41.0&40.9&1.43\\
     13856&2012-03-15 & Sgr A$^\star$&ACIS-S (HETG)&39.5&39.3&1.43\\
     13857&2012-03-17 & Sgr A$^\star$&ACIS-S (HETG)&39.0&39.0&1.43\\
     13854&2012-03-20 & Sgr A$^\star$&ACIS-S (HETG)&22.8&22.5&1.43\\
     14414&2012-03-23 & Sgr A$^\star$&ACIS-S (HETG)&19.8&19.8&1.43\\
     14427&2012-05-06 & Sgr A$^\star$&ACIS-S (HETG)&79.0&79.0&1.48\\
     13848&2012-05-09 & Sgr A$^\star$&ACIS-S (HETG)&96.9&96.9&1.48\\
     13849&2012-05-11 & Sgr A$^\star$&ACIS-S (HETG)&176.4&176.4&1.48\\
     13846&2012-05-16 & Sgr A$^\star$&ACIS-S (HETG)&55.5&54.0&1.48\\
     combined & -- & -- & -- & 705.5 & 702.7 & -- \\
     \hline
      \xmm\ &&&&&&\\
     ObsID & Start-Time & Target & pn-Filter & Exp & CL-Exp & $\theta_{off}$ \\
     & & & & (ks) & (ks) & (arcmin) \\
     \hline
     0554750401&2009-04-01 & Sgr A$^\star$&PFW-Medium&39.9&32.0&2.53\\
     0554750501&2009-04-03 & Sgr A$^\star$&PFW-Medium&44.3&38.1&2.57\\
     0554750601&2009-04-05 & Sgr A$^\star$&PFW-Medium&39.1&31.8&2.58\\
     0604300601&2011-03-28 & Sgr A$^\star$&PFW-Medium&48.8&27.3&2.52\\
     0604300801&2011-04-01 & Sgr A$^\star$&PFW-Medium&48.8&33.9&2.61\\
     0604300901&2011-04-03 & Sgr A$^\star$&PFW-Medium&46.9&18.8&2.58\\
     0604301001&2011-04-05 & Sgr A$^\star$&PFW-Medium&50.7&27.6&2.58\\
     0658600101&2011-08-31 & Sgr A$^\star$&PFW-Medium&49.9&48.1&2.04\\
     0658600201&2011-09-01 & Sgr A$^\star$&PFW-Medium&53.2&39.3&1.99\\
     0674600601&2012-03-13 & Sgr A$^\star$&PFW-Medium&21.5&8.4&2.50\\
     0674600701&2012-03-15 & Sgr A$^\star$&PFW-Medium&15.9&7.0&2.53\\
     0674601101&2012-03-17 & Sgr A$^\star$&PFW-Medium&28.0&10.6&2.53\\
     0674600801&2012-03-19 & Sgr A$^\star$&PFW-Medium&22.9&16.0&2.53\\
     0674601001&2012-03-21 & Sgr A$^\star$ &PFW-Medium&23.9&19.4&2.56\\
     combined & -- & -- & -- & 358.3 & 332.1& -- \\
     \hline
     \end{tabular}
 \end{minipage}
\end{table*}



\bsp	
\label{lastpage}
\end{document}